\documentclass[sigconf]{acmart}

\AtBeginDocument{%
  }

\setcopyright{acmcopyright}
\copyrightyear{2023} 
\acmYear{2023} 
\setcopyright{acmlicensed}\acmConference[SC-W 2023]{Workshops of The International Conference on High Performance Computing, Network, Storage, and Analysis}{November 12--17, 2023}{Denver, CO, USA}
\acmBooktitle{Workshops of The International Conference on High Performance Computing, Network, Storage, and Analysis (SC-W 2023), November 12--17, 2023, Denver, CO, USA}
\acmPrice{15.00}
\acmDOI{10.1145/3624062.3624181}
\acmISBN{979-8-4007-0785-8/23/11}

\usepackage{amsmath,amsfonts}
\usepackage{algorithm}
\usepackage{algpseudocode}
\usepackage{graphicx}
\usepackage{textcomp}
\usepackage{subcaption}
\usepackage{hyperref}
\hypersetup{
    colorlinks=true,
    linkcolor=blue,
    citecolor=red,
    filecolor=magenta,
    urlcolor=blue
    }

\usepackage{cleveref}
\usepackage{color}
\usepackage{xspace}
\usepackage{url}
\usepackage{lscape}
\usepackage{listings}
\usepackage{diagbox}
\usepackage[toc,page]{appendix}
\usepackage[inline]{enumitem}
\usepackage{stfloats}
\usepackage{balance}

\newcommand{\gko}{\textsc{Ginkgo}\xspace}

\definecolor{myhighlight}{RGB}{1,0,179}

\newcommand{\red}{\color{black}}

\lstdefinestyle{style1}{language=[11]C++,
    basicstyle=\ttfamily\footnotesize,
    commentstyle=\color{Green},
    frame=tb,
    keywords={typename,template,const, int,char,double,float,unsigned,void,bool,get, global},
    keywordstyle=\color{blue}\bfseries,
    ndkeywords={apply,kernel},
    ndkeywordstyle=\color{black}\bfseries,
    emph={
    gko,remove,complex,PrecType, ValueType, LogType, StopType,ToleranceType, BatchMatrixType, SimpleRelResidual,stop,BatchJacobi, BatchConvergence, BatchBicgstab, T},
    emphstyle={\color{ForestGreen}\bfseries},
    otherkeywords={*,>,<,.,;,-,!,=,~,:,_},
    morekeywords={*,>,<,.,;,-,!,=,~,:,_}
    }

\lstdefinestyle{style2}{language=[11]C++,
    basicstyle=\ttfamily\footnotesize,
    commentstyle=\color{Green},
    frame=tb,
    keywords={int,char,double,float,unsigned,void,bool,get, global},
    keywordstyle=\color{blue}\bfseries,
    ndkeywords={apply,kernel},
    ndkeywordstyle=\color{black}\bfseries,
    emph={
    gko,remove,complex,size_type, PrecType, ValueType, LogType, StopType,ToleranceType, BatchMatrixType, SimpleRelResidual,stop,BatchJacobi, BatchConvergence, BatchBicgstab, T},
    emphstyle={\color{ForestGreen}\bfseries},
    otherkeywords={*,>,<,.,;,-,!,=,~,:,_},
    morekeywords={*,>,<,.,;,-,!,=,~,:,_}
    }

\lstdefinestyle{style3}{language=[11]C++,
    basicstyle=\ttfamily\footnotesize,
    commentstyle=\color{Green},
    frame=tb,
    escapeinside={(*}{*)},
    keywords={auto,using,int,char,double,float,unsigned,void,bool,get,global,const,shared, ptr,std},
    keywordstyle=\color{blue}\bfseries,
    emph=[1]{T, ValueType, BatchBicgstab,ToleranceType, BatchConvergence},
    emphstyle={\color{ForestGreen}\bfseries},
    classoffset=0,
    otherkeywords={*,>,<,.,;,-,!,=,~,:,_},
    morekeywords={*,>,<,.,;,-,!,=,~,:,_}
    }

    \graphicspath{{./figs/}}

\newcommand{\batchdense}{\texttt{\textbf{BatchDense}}\xspace}
\newcommand{\batchcsr}{\texttt{\textbf{BatchCsr}}\xspace}
\newcommand{\batchell}{\texttt{\textbf{BatchEll}}\xspace}
\newcommand{\batchcg}{\texttt{\textbf{BatchCg}}\xspace}
\newcommand{\batchbicgstab}{\texttt{\textbf{BatchBicgstab}}\xspace}
\newcommand{\batchgmres}{\texttt{\textbf{BatchGmres}}\xspace}
\newcommand{\batchjacobi}{\texttt{\textbf{BatchJacobi}}\xspace}
\newcommand{\batchilu}{\texttt{\textbf{BatchIlu}}\xspace}
\newcommand{\batchisai}{\texttt{\textbf{BatchIsai}}\xspace}
\newcommand{\batchtrsv}{\texttt{\textbf{BatchTrsv}}\xspace}

\usepackage[acronym,xindy,toc]{glossaries} 
\makeglossaries

\newacronym{hpc}{HPC}{High-performance Computing}

\newacronym{gemm}{GEMM}{General Matrix Multiplication}

\newacronym{avx}{AVX}{Advanced Vector Extensions}

\newacronym{fp}{FP}{Floating-point}

\newacronym{fma}{FMA}{Fused Multiply-add}

\newacronym{jit}{JIT}{Just-In-Time}

\newacronym{mkl}{oneMKL}{Intel oneAPI Math Kernel Library}
\newacronym{dpl}{oneDPL}{Intel oneAPI DPC++ Library}

\newacronym{gpu}{GPU}{Graphics Processing Unit}

\newacronym{cpu}{CPU}{Central Processing Unit}

\newacronym{dpc++}{DPC++}{Data Parallel C++}

\newacronym{pvc}{Ponte Vecchio GPU}{Intel GPU Max 1550}

\newacronym{smt}{SMT}{Simultaneously Multithreaded}

\newacronym{alu}{ALU}{Arithmetic Logic Unit}

\newacronym{slm}{SLM}{Shared Local Memory}

\newacronym{simd}{SIMD}{Single Instruction Multiple Data}

\newacronym{grf}{GRF}{General Register File}

\newacronym{gr}{GR}{General-purpose Register}

\newacronym{xve}{XVE}{$\mathrm{X}^\mathrm{e}$ Vector Engine}

\newcommand{\xe}{$\mathrm{X}^\mathrm{e}$}

\newacronym{xmx}{XMX}{$\mathrm{X}^\mathrm{e}$ Matrix Engine}

\newacronym{xc}{XC}{$\mathrm{X}^\mathrm{e}$-Core}

\newacronym{emib}{EMIB}{Embedded Multi-die Interconnect Bridge}

\newacronym{fpga}{FPGA}{Field Programmable Gate Array}

\newacronym{ndrange}{ND-Range}{N-dimensional range}

\newacronym{hbm}{HBM}{High Bandwidth Memory}

\begin{document}

\title{Porting Batched Iterative Solvers onto Intel GPUs with SYCL}

\author{Phuong Nguyen}
\email{phuong.nguyen@icl.utk.edu}
\affiliation{%
  \institution{University of Tennessee, Knoxville}
  \country{USA}
}
\author{Pratik Nayak}
\email{nayak@kit.edu}
\affiliation{%
  \institution{Karlsruhe Institute of Technology}
  \country{Germany}
}

\author{Hartwig Anzt}
\email{hanzt@icl.utk.edu}
\affiliation{%
  \institution{University of Tennessee, Knoxville}
  \country{USA}
}

\begin{abstract}
  Batched linear solvers play a vital role in computational sciences,
  especially in the fields of plasma physics and combustion
  simulations. 
  With the imminent deployment of the Aurora Supercomputer and other upcoming
  systems equipped with Intel GPUs, there is a compelling demand to expand the
  capabilities of these solvers for Intel GPU architectures.

  In this paper, we present our efforts in porting and optimizing the batched
  iterative solvers on Intel GPUs using the SYCL programming model.
  {\red
  These new solvers achieve impressive performance on the  \glspl{pvc}
  which surpass our previous CUDA implementation on NVIDIA H100 GPUs by
  an average of 2.4x for the PeleLM application inputs. 
  }
  The batched solvers are ready for production use in real-world scientific
  applications through the Ginkgo library{\red, complementing the performance
  portability of the batched functionality of Ginkgo. 
  }
\end{abstract}

\ccsdesc[500]{Mathematics of computing~Solvers}
\ccsdesc[500]{Mathematics of computing~Mathematical software performance}
\ccsdesc[500]{Computing methodologies~Massively parallel algorithms}
\keywords{SYCL, Performance Portability, Batched Linear Solvers, Intel GPUs}



\maketitle

\section{Introduction}
\label{sec:intro}

Batched iterative solvers have recently received a lot of attention due to their
efficiency in solving batches of small and medium-sized sparse
problems{\red\cite{Kim2023,Anzt2016,kashiBatchedSparseIterative2022}}.
Similar to the monolithic problems, batched iterative solvers in particular
outperform their direct counterparts if they can use the solution of a similar
problem, for example, the previous system in a Picard loop, as the initial guess,
which can dramatically shorten the iteration process. For a sequence of linear systems, direct solvers always have
to start from scratch with a complete (sparse) factorization for each problem. Generally, sparse direct solvers have the
disadvantage that the fill-in in the factorization process is unknown a priori. In the batched case, this conflicts with the goal of packing all operations of the solve into a
single kernel to reduce the main memory access. So batched sparse direct solvers
will typically compose of two kernels with a memory allocation in-between, while
batched iterative solvers can execute as a single kernel that can leverage data
locality.

Batched routines were originally developed for NVIDIA GPUs, as these GPUs were
the first to be used in scientific
computing{\red~\cite{10.1145/3148226.3148230,
DBLP:conf/supercomputer/AbdelfattahHTD16,carrollBatchedGPUMethodology2021a}}.
The batched kernels were written in NVIDIA's CUDA programming
ecosystem{\red ~\cite{10.1145/1401132.1401152}}. 
In the last years, an increasing number of leadership systems are equipped with GPUs
from other vendors, including AMD and Intel, there is a need to port these
routines beyond the CUDA backend. 
SYCL drew our interests due to its performance efficiency and portability
on different architectures, such as CPUs, GPUs, and
FPGAs~{\red\cite{9309052,8945798}}. 
SYCL is a cross-platform abstraction layer inspired by OpenCL{\red
~\cite{10.1145/3388333.3388649}}. Its underlying
fundamental principles of portability and efficiency enable the composition of
code, in a "single-source" style using completely standard C++, for 
heterogeneous architectures.  SYCL's increasing
popularity has led to the development of a diverse range of implementations
within its
ecosystem~{\red\cite{10.1145/3585341.3585343,10.1145/3388333.3388658}}. Among them, the Intel oneAPI Toolkit provides 
a SYCL implementation and compilers which are highly efficient for Intel
GPUs~{\red \cite{oneapi2023}}.

Historically, batched functionality was developed for scenarios where many small
and independent problems had to be tackled in parallel with the same algorithm,
with each small problem being too small to fully use the available compute
resources. In that sense, batched functionality is suitable for data-parallel
problems. Typical use cases for batched functionality are parallel applications
of a linear operator as a dense or sparse batched matrix-vector
multiplication~\cite{batchedgemv}, the parallel solution of a set of
pairwise-independent linear systems~\cite{batchedlu}, or the parallel singular
value decomposition~\cite{10.1016/j.parco.2017.09.001}. The use of these methods
spans from high-order FEM schemes over tensor contractions in quantum Hall
effects, astrophysics, metabolic networks, and quantum chemistry to image and
signal processing. With the rise of machine learning and the heavy use of deep
neural networks, the batched dense matrix-matrix
multiplication~\cite{DBLP:conf/supercomputer/AbdelfattahHTD16} has become the
most prominent use case for batched functionality. In many cases, the
data-parallel problems arise by breaking down a large problem into many small
problems that can be handled more efficiently if they are considered
independently. A colorful example is the application of a block-Jacobi
preconditioner that can be expressed either by applying a block-diagonal matrix
to a global vector or by applying a set of small dense matrices to vector
segments, with the latter allowing for a more localized kernel execution. In consequence, batched operations are traditionally designed to perform
the same pre-defined sequence of operations on all problems of the set. This
allows for handling all problems in a SIMD-fashion: the problem-individual
properties have no influence on the batched kernel execution. Given the strong
demand for batched functionality and the possibility to leverage hardware
characteristics in the performance optimization of batched functionality, the
community has agreed on a de-facto batched BLAS interface
convention~\cite{10.1145/3431921} that mostly adheres to the vendor
implementation of batched BLAS libraries~\cite{cuBLAS, MKL}.

Batched dense functionality is often used in sparse linear algebra computations,
for example when handling problems that contain small dense blocks, such as
high-order finite element discretizations or supernodal factorizations. At the
same time, there exists little work on batched sparse functionality. One of the
earliest efforts on batched sparse functionality is the batched sparse matrix
vector product kernel developed by Collins et
al.~\cite{10.1145/3148226.3148230}. However, the design and execution of the
batched sparse matrix vector multiplication kernel is different from the
traditional batched functionality as it uses different storage formats for the
distinct sparse problems and launches suitable kernels -- matching the storage
format -- via runtime polymorphism.
With regard to batched sparse direct solvers, we mention in passing some work on
batched tri-diagonal and penta-diagonal systems
\cite{cuThomasBatch,carroll2021batched}. However, these direct methods are
restricted to batched tri- and penta-diagonal systems and typically utilize
only one GPU thread per batch entry to solve it sequentially. While this
approach is advantageous for certain types of problems, it does not utilize the
fine-grained parallelism as is possible with iterative methods.

Recently, there have been some developments in the batched iterative solver, as an
alternative to batched direct
methods~\cite{aggarwalBatchedSparseIterative2021,liegeoisPerformancePortableBatched2023}.
In particular, for GPUs, it has been shown that the batched iterative
methods can match/outperform the batched direct counterparts. Applications such
as combustion and fusion plasma simulations need to solve hundreds of thousands
of small to medium linear systems, each sharing a sparsity pattern. For the linear
solution of these systems, placed inside a non-linear loop, it is advantageous
to use an iterative solver, as that allows to incorporate an initial guess which
can accelerate the linear system solution within the outer loop. Additionally,
we might not need to solve the system to machine precision accuracy but can
control the solution accuracy based on the parameters of the outer non-linear
loop~\cite{aggarwalBatchedSparseIterative2021}.

In this paper, we extend our previous GPU-native batched linear solvers to the SYCL
ecosystem and showcase the performance on Intel GPUs. We compare the performance
of the batched solvers on Intel GPUs against the NVIDIA GPUs, both on their
vendor native programming models (SYCL and CUDA respectively). We explore the
subtleties in the kernels and the differences required to optimize performance
for both programming models. Focusing on a simple 3-point stencil discretization problem, we
study scaling and demonstrate that we outperform the state-of-art for all problem
sizes. We also use matrices derived from the
Pele reaction flow simulation application that uses
SUNDIALS~\cite{hindmarsh2005Sundials} to solve the ODE linear
systems, which lend themselves to batched solutions.

In \Cref{sec:background}, we provide some background on batched solvers, their
need in applications, and aspects that need to be considered for batched
iterative solvers. We also briefly detail the SYCL programming model and its
features and the Intel GPU hardware characteristics. In \Cref{sec:impl}, we
detail the design of our batched sparse iterative solvers. In particular, we
showcase our batched iterative solver design that comes with the flexibility of
using different preconditioners, stopping criteria, and sparse matrix storage
formats, and monitor the solver convergence for each system in the batch
individually. We also elaborate on the specific SYCL optimizations that enable
us to maximize performance on the Intel GPU.

After presenting all implementation details, in \Cref{sec:evaluation} we
investigate the hardware performance, and
time-to-solution of the batched sparse iterative solver technology. This
includes the evaluation for benchmark problems arising in real-world PeleLM
combustion applications. {\red We also briefly discuss the performance
portability and productivity of the implemented SYCL-based solvers. }
In \Cref{sec:conclusion}, we summarize our findings and
provide a roadmap for the extension of the batched sparse iterative
functionality that we plan to provide in the Ginkgo open-source library
\cite{ginkgo-joss, ginkgo-arxiv}.

In summary, we make the following contributions:
\begin{itemize}
  \item Successfully porting the batched iterative solvers onto the Intel GPUs
        using the SYCL programming model.
  \item Performance tuning for the ported solvers on the Intel GPUs for a
        wide range of matrix sizes.
  \item Performance evaluation for two different paradigms: a three-point
        stencil matrix used to study scaling behaviour, and matrices from the
        PeleLM application.
  \item A thorough evaluation of the performance of the batched iterative
        solvers on Intel GPUs and their comparison against the latest NVIDIA H100
        GPU, both using the vendor native programming models.
\end{itemize}

\section{Background}
\label{sec:background}

Consider applications such as a chemical reaction or astrophysical simulations, which aim to
evolve the reactive flow in time on a discretized 3D mesh. These simulations
typically operator split the reactions from the hydrodynamics and hence require
a solution of many independent chemical reaction ODEs. The resulting chemical reaction
equations are usually very stiff, requiring the usage of implicit time stepping
schemes such as the Backward Differentiation
Formula(BDF)~\cite{hindmarsh2005Sundials}. In each time step, one
needs to solve a non-linear system. Solving this non-linear system with, for
example, a Newton iteration requires the solution of linear systems. These linear
systems characterize the reaction of species in the domain. As the species in
the domain are the same across all cells and the reaction matrix is defined for
the species, for each spatial discretization cell, we need to solve a linear
system, with the linear system for all cells sharing the sparsity pattern.

Many other applications such as fusion plasma simulations or finite element
simulations also require the solution of independent linear
systems~\cite{kashiBatchedSparseIterative2022},
and particularly within a non-linear iteration loop.

\subsection{Batched iterative solvers}
\label{sec:background:batched}

In contrast to batched direct solvers, batched iterative solvers provide the
possibility to vary the solution accuracy, which can be beneficial to reduce the runtime of the non-linear iteration. Additionally, iterative solvers
can incorporate solution information in the form of an initial guess, which can
accelerate the overall time to solution.

This tunable accuracy and initial guess capabilities come at the cost of
complexities in design and implementation. Iterative solvers do not follow a
pre-defined execution, but the number of iterations depends on the matrix
properties, the stopping criterion, and the precision format being used. Due to the
nature of iterative solvers, the design needs to take into account the efficient
composition of kernels with the plethora of parameters that are necessary for
an efficient iterative solution.

With hierarchical architecture such as GPUs, the design needs to minimize memory
movement, maximize local memory usage, and the overall occupancy of the GPU to
maximize the performance. This optimization typically has to account for the
target problem or the target problem class. We here focus on linear systems of
small to medium size (of the order of 10 to 2,000 rows) and system matrices
sharing the same sparsity pattern.

\subsection{Overview of \glspl{pvc} architecture}
\label{sec:background:intelgpus}
To motivate the design of batched iterative solvers for Intel GPUs, we provide some technical background on the Intel GPU architecture.

The \xe-core is the smallest thread-level building block of the PVC GPU. Each
\xe-core consists of eight \glspl{xve}, each of which have a 512-bit register,
and can therefore perform 256 FP32 or FP64 operations per cycle.  
Each \xe-core can execute eight multithreads simultaneously, with each
hardware thread having 4096 bytes of private memory in the form of 128 general purpose
registers.

Each PVC GPU consists of 2 stacks which are connected via a \textit{Stack-to-Stack} link.
Each stack has its own dedicated
resources: 4 \xe-slices, a Level 2 Cache, and a directly connected 64 GB of \gls{hbm}.
Each \xe-slice consists of 16 \xe-cores. 
Overall, each PVC GPU has 128 \xe-cores and 128 GB of \gls{hbm}.

Even though the two stacks have separately connected \glspl{hbm}, this
\glspl{hbm} can be accessed directly by the other stack. This feature
enables fast and efficient communication between the two stacks, via the
\gls{hbm}.

In practice, when running an application, this two-stack GPU can be seen as a
single GPU device. The GPU driver and the runtime work together to automatically
distribute the workloads across the two stacks. This describes the so-called \textit{implicit
scaling mode}. In contrast to the \textit{implicit scaling mode}, the
\textit{explicit scaling mode} allows users to explicitly allocate the
workloads and memory placement for each stack. Each stack then executes its own
workloads.

As one may find more familiar with NVIDIA GPU terminology, the mapping between
NVIDIA's terminology and Intel's terminology is provided
in~\Cref{tab:background:gpu_map}. Overall, with the exception that the PVC GPUs
consist of 2 stacks, all other terminologies can be directly paired with the
NVIDIA terminology.

\label{sec:background:sycl}
\begin{figure}[t]
  \includegraphics[width=\linewidth]{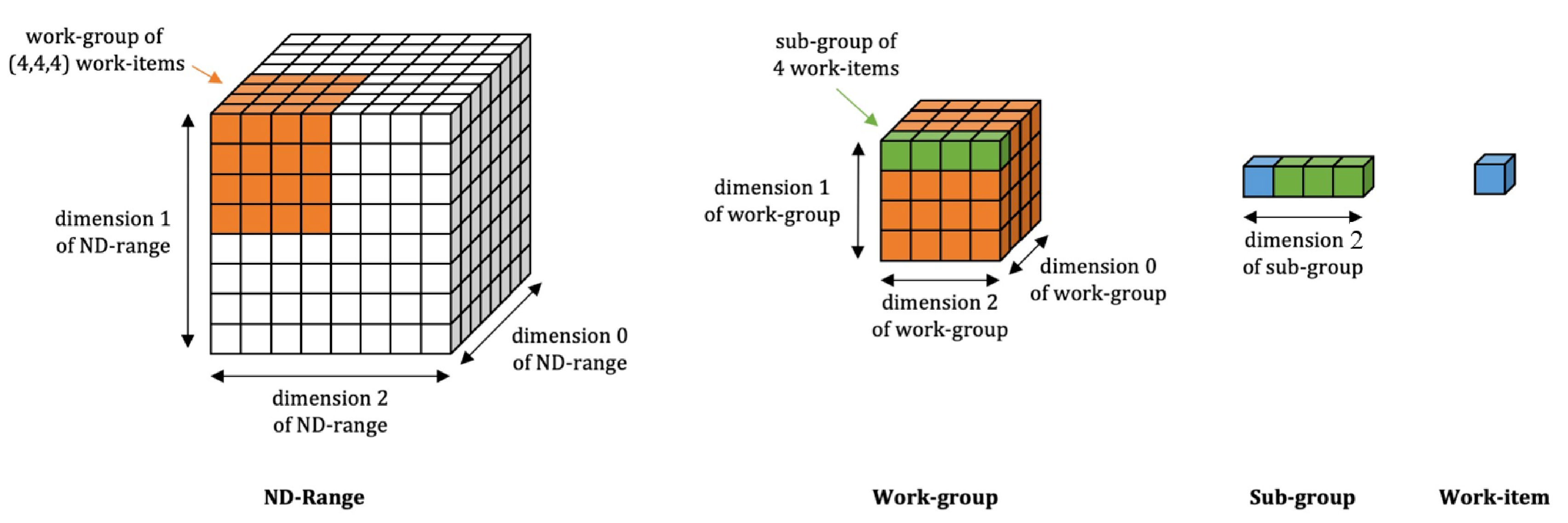}
  \caption{Hierarchy of the SYCL kernel index space~\cite{dpcpp2021}.} 
  \label{fig:background:nd_range}
\end{figure}

\begin{table}[t]
  \caption{GPU architecture terminology mapping~\cite{oneapi2023}}
  \label{tab:background:gpu_map}
  \begin{tabular}{ l l }
    \toprule
    CUDA Capable GPUs & \glspl{pvc} \\ 
    \midrule
    CUDA Core &  \gls{xve} \\
    Streaming Multiprocessor & \gls{xc}  \\
    Processor Cluster & \xe-Slice \\
    N/A & \xe-Stack \\
    \bottomrule
  \end{tabular}
\end{table}

\begin{table}[t]
  \caption{Execution model mapping from CUDA to SYCL~\cite{oneapi2023}}
  \label{tab:exec-model}
  \begin{tabular}{ll}
    \toprule
    CUDA & SYCL \\
    \midrule
    Thread & work-item \\
    Warp & sub-group \\
    Block & work-group \\
    Grid & ND-range \\
    \bottomrule
  \end{tabular}
\end{table}

\subsection{SYCL Programming Model}
SYCL is a Khronos Group language standard that enables developers to express
data-parallel computations using standard C++ templates and lambda functions,
abstracting the underlying hardware complexity and allowing seamless execution
on diverse accelerators. 

The SYCL kernel consists of the main kernel computation which is expressed as a
C++ lambda function, the argument values associated with the kernel, and the
parameters that define an index space.
The kernel index space is often defined via an ND-Range.

\Cref{fig:background:nd_range} illustrates the index hierarchy of
the kernel instance, in which the smallest kernel execution unit is called a
\textit{work-item}. Multiple consecutive work-items can be organized into a 1-dimensional
set called a \textit{sub-group}. The computation of a sub-group can be processed by
one or few SIMD operations on a \gls{xve}. Additionally, one can also perform
collective operations such as broadcast, shuffle, reduction, etc within a sub-group.

A \textit{work-group} consists of a 1-,2-, or {\red 3-}dimensional set of consecutive sub-groups. Each
work-group has a local memory which is shared among all the work-items
within a work-group. Depending on the implementation and the hardware
availability, this \gls{slm} can be mapped into different physical memories.
On Intel GPUs, the \gls{slm} is allocated on the L1 cache.
Typically, the work-items in a work-group are executed together on a \xe-core.
Depending on the work-group size and availability of the L1 cache, each \xe-core
can handle multiple work-groups at a time.
Additionally, the work-items within a work-group can be synchronized via local memory fences but
synchronization across different work-groups is not possible in SYCL.

As one may be more familiar with CUDA terminologies, the execution model mapping between CUDA
and SYCL can be found in Table~\ref{tab:exec-model}.

\section{Implementing batched iterative solvers}
\label{sec:impl}

In this section, we discuss the design and the interface of batched sparse
iterative solvers. Our goal is to develop batched sparse iterative
solvers to be flexible in terms of accepting a preconditioner transforming the
linear systems $A_ix_i=b_i, i=1\dots n$ into the preconditioned system
$M_iA_ix_i = M_ib_i, i=1\dots n$ with $M_i$ being a preconditioner adjusted to
the specific system $A_i$, but all preconditioners $M_i$ being of the same
preconditioner type.

The objective of a batched solver interface is to leverage the data parallelism
present in the batched problem at hand and map it to the hardware parallelism
available. Batched solvers are favorable in cases where the individual matrices
are relatively small, and a
large numbers of these independent linear systems needs to be solved. The criteria that
influence the design and implementation are the following:

\begin{enumerate}
    \item The size of the individual batch entries: the number of rows and the number of non-zeros.
    \item The number of linear systems to be solved.
    \item Common sparsity patterns between the batched matrices, if any.
    \item Properties of the batched linear systems that influence convergence (condition numbers, etc.)
\end{enumerate}


To enable support for a wide variety of applications, Ginkgo supports different
batched matrix formats, solvers, and preconditioners, as shown in
\Cref{tab:batched-features}. We note that due to the templated design, any of the columns
can be combined with another, with only a few exceptions (such as \batchisai
needing the \batchcsr matrix format).

\begin{table}[t]
\caption{Batched feature support in Ginkgo }
\label{tab:batched-features}
\centering
 \begin{tabular}{llll}
   \toprule
 Mat. formats & Solvers & Preconditioners & Stop. criteria\\
   \midrule
 \batchdense & \batchcg & \batchjacobi  & Absolute\\
 \batchcsr & \batchbicgstab & \batchilu &  Relative\\
 \batchell & \batchgmres & \batchisai &\\
 & \batchtrsv& &\\
   \bottomrule
\end{tabular}
\end{table}

\subsection{Batched matrix formats}

Sparse matrices typically store an array of non-zero values, as well as integer
arrays encoding the sparsity pattern. For our problem space, all the matrices in
the batch share the same sparsity pattern. Therefore, to minimize memory
requirements, we store only one copy of the sparsity pattern for the batched
matrix formats. We additionally implement two batch matrix formats, one general format,
\batchcsr, and a specialized format, \batchell in addition to the dense matrix
format, \batchdense.

The \batchcsr matrix format is based on the popular Compressed Sparse Row (CSR) matrix
storage format, where one stores an array of column indexes per row
corresponding to each non-zero value in the matrix. An accumulated sum of the
number of non-zeros per row is additionally necessary. This matrix format is
suitable for general matrices with large variations in the number of non-zeros
per row and performs generally well for most sparsity patterns. The \batchcsr is an
extension of this format where we store the column indexes and the row pointers
for only one matrix and store the values of all the matrices.

For matrices that have a similar number of non-zeros in every row, we can
optimize the storage by padding the rows to a uniform number of non-zeros per
row, removing the need for a pointers array. This also gives us additional
advantages in terms of coalesced accesses. The \batchell matrix format stores
one set of column indexes and the values of all the batch entries. In contrast
to \batchcsr, we store the column indexes and the values in column-major
allowing for coalesced accesses which is suitable for GPUs.

\begin{figure}
    \centering
    \includegraphics[width=0.775\columnwidth]{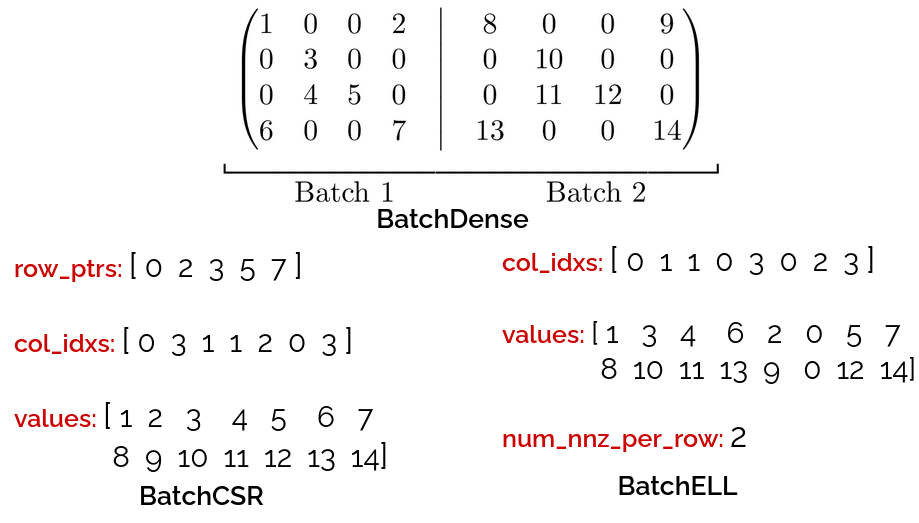}
    \caption{Batch Matrix Storage formats - \batchdense, \batchcsr and \batchell}
    \label{fig:storage_formats}
\end{figure}

\Cref{fig:storage_formats} visualizes the schematic and the storage
requirements of \batchcsr and \batchell compared to the \batchdense format. With
batched sparse matrix formats, the additional cost of storing the indexes and
the pointers can be easily amortized over an increasing number of systems in the
batch. The storage requirements therefore are:

\begin{enumerate}
    \item \batchdense: \texttt{num\_matrices x num\_nnz\_per\_matrix}
    \item \batchcsr: \texttt{[num\_matrices x num\_nnz\_per\_matrix] \\
+ [(num\_rows + 1) x 1]
+ [num\_nonzeros\_per\_matrix x 1]
}
    \item \batchell: \texttt{[num\_matrices x num\_nnz\_per\_matrix]\\
+ [num\_nnz\_per\_row x num\_rows x 1]}

\end{enumerate}

\subsection{Batched solver kernels}

Iterative solvers such as CG, BiCGSTAB, GMRES can be easily composed of BLAS 1, BLAS 2, and
sparse matrix vector operations~\cite{saadIterativeMethodsSparse2003}. Ginkgo
supports different batched versions of the iterative solvers, suitable for
matrices with different properties and these are listed in
\Cref{tab:batched-features}. With our problem space consisting of small to
medium-sized linear systems, and our aim to optimize the compute and memory
usage, we map one work-group to one linear system. This enables us to write
efficient kernels for each linear system without worrying about global
synchronization between workgroups (as each linear system is independent and
requires no communication).

With the sparse matrix vector product being the workhorse of the Krylov solvers,
we implement tuned SpMV kernels for each batched matrix format.
For the \batchcsr matrix, we implement a
sub-group to row-based mapping for matrices which provides good performance
for general matrices. For matrices that are more balanced and have only a few
nonzeros per row, the \batchell matrix format is more suitable, which handles
one row per work item removing the need to communicate between
thread using sub-warp reductions~\cite{kashiBatchedSparseIterative2022}.

In addition to the SpMV kernel, kernels such as dot, scalar addition, and norm
are also implemented. Reduction operations such as dot and norm are
implemented using the reduction over the whole work-group which is a primitive
function provided by SYCL. For small matrices, it is more efficient to
implement the reduction within a subgroup since we do not need to
read/write through the \gls{slm}.
These reduction operations were implemented in a different fashion compared to our CUDA-based
solvers as in CUDA only warp-level reductions are used as no efficient
thread-block level reduction operations are available.

We note that these building blocks are all device kernels and inlined, involving
no host-device transfers. This enables the compiler to
optimize the entire solver kernel as a whole. Additionally, the SpMV, scalar
operations, dot, and norm kernels are shared between the different solvers,
reducing code duplication and improving code sustainability.

\subsection{Multi-level dispatch mechanism}

\begin{figure}[t]
 \centering
 \includegraphics[width=0.95\linewidth]{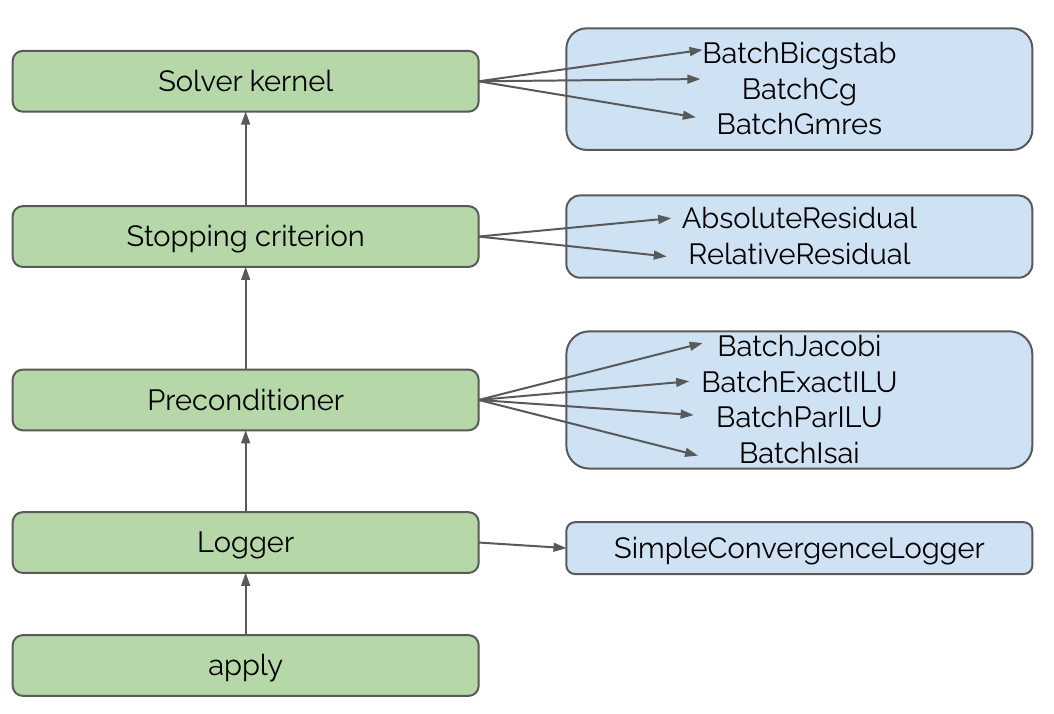}
 \caption{Multi-level dispatch mechanism}
 \label{fig:impl:dispatch}
\end{figure}

We design a multi-level dispatch mechanism as shown in \Cref{fig:impl:dispatch} that preserves flexibility, enabling
the runtime choice between the different matrix formats, solvers, stopping criteria
and preconditioners.

\subsection{Minimizing kernel launch latency}

For batched solvers, the time to solution for one batch item can be very small,
particularly for small linear systems. Launching one kernel for each batch item
or for even a few items at once is therefore intractable. To minimize kernel
launch overhead, we gather all functionality in a single kernel handling all
items of the batch.

The strategy of packing all functionality into a single kernel enables the
compiler to optimize the templated kernel as a single instance after the
multi-level dispatch mechanism has instantiated the different kernel options
(precision format, matrix format, preconditioner, stopping criterion). The
modern C++ templating mechanism in this case not only avoids code complexity,
but also kernel branching which is prohibitively expensive when handling small
problems.

The kernel execution then handles the solution of all items in the batch with the same solver configuration.

\subsection{Maximizing local memory usage}
\label{sec:impl:max_mem_usage}
Maximizing the performance of the batched solvers requires efficient usage of
both the compute and memory hierarchy. 
It is essential to keep frequently used data in \gls{slm} since
accessing this memory has a lower latency than the global memory. 

In our implementation, we map one linear system into one work-group, i.e. each
work-group solves one linear system at a time. Within that, each work-group keeps
its intermediate vectors which requires for the iterative solvers to solve the
system separately. The sizes of these vectors depends on the size of the batch
item matrix.
To reduce the latency of memory accesses, it is
beneficial to allocate these intermediate vectors on the \gls{slm}.
The pre-conditioned matrix and a copy of the result vector \textit{x}
are also allocated on the \gls{slm} as they are repeatedly used in
the solvers kernel. Besides, the system matrix and the right-hand side are
read-only data that needed to fetched from the global memory in every iteration,
but their sizes are relatively large for being kept inside the \gls{slm}.
Therefore, caching these data into another level cache, for example, L2, is
favorable.

For medium to large matrix sizes, allocating all these objects on the \gls{slm}
is impossible as the size of the \gls{slm} is limited.
For each batched iterative solver type, we prioritize these intermediate vectors based on its
usage frequency and sizes. 
Based on this priority, the solvers dynamically
determine at runtime how many vectors can be allocated on the \gls{slm},
given the input matrix size and the available \gls{slm} memory on the device. 
{\red The host then selects and dispatches the appropriate kernel which
allocates the required amount of \gls{slm} and assign these objects accordingly.}

\def\spmvcol{black}
\def\constcol{black}
\def\othercol{black}
\newcommand{\spmvvec}[1]{{\color{\spmvcol} \boldsymbol{#1}}}
\newcommand{\constvec}[1]{{\color{\constcol} \boldsymbol{#1}}}
\newcommand{\ivec}[1]{{\color{\othercol} #1}}
\newcommand{\bld}[1]{\boldsymbol{\ivec{#1}}}

\begin{algorithm}[t]
\begin{algorithmic}[1]
\For{$b < N_{batches}$}
\State $\spmvvec{r} \gets \constvec{b}-\constvec{A}\bld{x}, \bld{z} \gets
\constvec{M}\spmvvec{r},
\bld{p} \gets \bld{z},  \spmvvec{t} \gets \mathbf{0}$
\State $\rho \gets \spmvvec{r}\cdot\bld{z}, \alpha \gets 1, \hat{\rho} \gets 1$
\For{$i < N_{iter}$}
    \If{$ \lvert \rho \rvert < \tau$}
        \State \texttt{break}
    \EndIf
    \State $\spmvvec{t} \gets \constvec{A} \bld{p}$
    \State $\alpha \gets \frac{\rho}{\bld{p}\cdot \spmvvec{t}}$
    \State $\bld{x} \gets \bld{x} + \alpha \bld{p}$
    \State $\spmvvec{r} \gets \spmvvec{r} - \alpha \spmvvec{t}$
    \State $\bld{z} \gets $ \Call{precond}{$\spmvvec{r}$}
    \State $\hat{\rho} \gets \spmvvec{r}\cdot\bld{z}$
    \State $\bld{p} \gets \bld{z} +  \frac{\hat{\rho}}{\rho} \cdot \bld{p}$
    \State $\rho \gets \hat{\rho}$
\EndFor
\EndFor
\end{algorithmic}
\caption{The \batchcg solver.}\label{alg:cg}
\end{algorithm}

The priority for storing those objects on the \gls{slm} can be illustrated
through the \batchcg, for example. Its algorithm can be found in
\Cref{alg:cg}. Based on the usage frequency and the size of these
objects, the priority we assign in decreasing order is:
$r$, $z$, $p$, $t$, $x$. 
The preconditioner workspace is also allocated on the \gls{slm} if the \gls{slm}
is still available.

\subsection{Optimizations based on the matrix size}
\label{sec:impl:optimizations}
It is important for the batched solvers to have good performance across a
wide range of matrix sizes so that it meets the needs of different real-world
applications. In this section, we discuss the optimizations techniques we used
for our SYCL-based batched solvers relying on the size of the input matrices.

The performance of GPU kernels often depends highly on the execution
configuration.
Since we assign one linear matrix system to one work-group, it is beneficial to
select the work-group size based on the input matrix size.
In our implementation, the work-group size is chosen dynamically at runtime
depending on the number of rows of the input matrix, as follows:
\begin{itemize}
  \item {\red The work-group size should not exceed the maximal work-group size supported by the
    device.}
  \item {\red The work-group size should be divisible by the sub-group size.}
  \item The work-group size should be at least equal to the number of rows so
    that the \textit{SpMV} kernel performs efficiently.
\end{itemize}
Thus, for small matrices, in the case when the number of rows is divisible by the sub-group size,
the number of rows is chosen as the work-group size. Otherwise, we choose the
work-group size equals to the next round-up number by sub-group size of the number of
rows. {\red This round-up strategy shows a performance improvement for some input
cases, as shown in Table~\ref{tab:appendix:roundup} (Appendix~\ref{appendix:performance}).
The maximal work-group size is selected in the case of large input matrix
cases.}
These selecting work-group strategies not only increases the thread utilization within a
work-group but also increases the number of possible work-groups that can be scheduled on
the same \gls{xve} at a time, enhancing the GPU occupancy.

Besides the work-group size, the sub-group size also plays an important role in
the performance of the GPU kernels in SYCL. In fact, the Intel PVC GPUs support two
different sub-group sizes: 16 and 32.
For our batched solver implementations, empirically, we measured that
a sub-group size of 16 is better for the small matrices, while a size of
32 performed better for larger input matrices. Thus, the SYCL-based batched solvers are
implemented in such a way that they can choose the sub-group size, in addition to
the work-group size, dynamically, at runtime. In practice, since SYCL
only allows to enforce the sub-group size with a compile-time known number, we
use \textit{C++ templating} to instantiate kernels with all possible
sub-group size values and the proper kernel is selected at runtime based on the
matrix size.

Additionally, as the matrix size also dictates whether the reduction operations
use a single sub-group or the whole work-group,
we implemented our batched solvers in such a way that the selection can happen
at runtime for these reduction operations as well.
Again, we use the \textit{C++ templating} ideas to minimize runtime overhead.

Overall, due to these optimizations, the batched solvers switch between
different paths in the kernel launch stage, and the selected kernel depends on the matrix size. Since the
thresholds between small and large matrix sizes are different for different GPUs
capabilities, these thresholds need to be determined experimentally for each
targeted device before using these solvers to ensure optimal performance for
that architecture.

\section{Performance evaluation}
\label{sec:evaluation}

In this section, we evaluate runtime and scalability of the batched solvers on
the latest Intel GPUs with SYCL and benchmark the performance against the batched solver
implementation for NVIDIA GPUs with CUDA.

\subsection{Experimental setup}
\label{sec:setup}
To evaluate our SYCL-based batched iterative solvers, we consider two classes of
inputs, summarized in \Cref{tab:measurement:inputs}:

\begin{itemize}
  \item Matrices generated from a 3-point stencil, whose number of rows can
        be increased as necessary to study the scaling behaviour of the batched
        solvers.
 \item Matrices from the PeleLM+SUNDIALS application, which consists of
        matrices derived from different reactive flow simulations. Each
        mechanism gives us a different matrix set and we have as many items in
        the batch as the number of cells in the mesh. As we would like to have a smaller test case, we extract the matrices from the application for a
        few cells and replicate it to emulate the solution for a larger mesh.
        The matrices for this application are fairly small (22 rows to 144 rows)
        and relatively dense. More information on this dataset is available in~\cite{aggarwalBatchedSparseIterative2021}.
\end{itemize}
In the experiment, all input matrices are stored in the \batchcsr format. We
study the performance for two batched iterative solvers, the \batchcg and the
\batchbicgstab. Additionally, the PeleLM+SUNDIALS matrices use a scalar Jacobi
preconditioner to accelerate convergence.

\begin{table}[t]
  \caption{Reference for data inputs}
  \label{tab:measurement:inputs}
  \begin{tabular}{l ccc}
    \toprule
    Input case & \# Unique matrices & Matrix size & \# Nnz/matrix \\
    \midrule
    \textit{3pt stencil} & - & -  & 3 x $n_{rows}$\\
    \textit{drm19} & 67 & 22 x 22 & 438 \\
    \textit{gri12} & 73 & 33 x 33 & 978 \\
    \textit{gri30} & 90 & 54 x 54 & 2560 \\
    \textit{dodecane\_lu} & 78 & 54 x 54 & 2332 \\
    \textit{isooctane} & 72 & 144 x 144 & 6135 \\
    \bottomrule
  \end{tabular}
\end{table}

\begin{table}[t]
  \caption{GPUs specifications}
  \label{tab:measurement:gpus}
  \begin{tabular}{ l  c c c c}
    \toprule
    & A100  & H100 & PVC-1S & PVC-2S\\
    \midrule
    FP64 Peak (TFLOPs) & 9.7 & 26 & 22.9 & 45.8 \\
    HBM BW Peak (TB/s) & 1.6 & 2.0  & 1.6 & 3.2 \\
    Shared Local Mem. (KB) & 192 & 228 & 128 & 128 \\
    \bottomrule
  \end{tabular}
\end{table}

We compare the batched iterative solvers on three different GPUs:
\begin{enumerate}
  \item NVIDIA A100 80GB PCIe using CUDA 11.8.0
  \item NVIDIA H100 PCIe Gen 5 using CUDA 11.8.0
  \item Intel Data Center GPU Max 1550 (PVC) using Intel(R) oneAPI DPC++/C++ Compiler 2023.2.0
\end{enumerate}

Some salient features of the GPUs that are relevant in our case is tabulated in \Cref{tab:measurement:gpus}.


\subsection{Scaling evaluation with the synthetic input}
\begin{figure}[t]
  \centering
\hfill
\begin{subfigure}{\linewidth}
 \centering
  \includegraphics[width=0.65\linewidth]{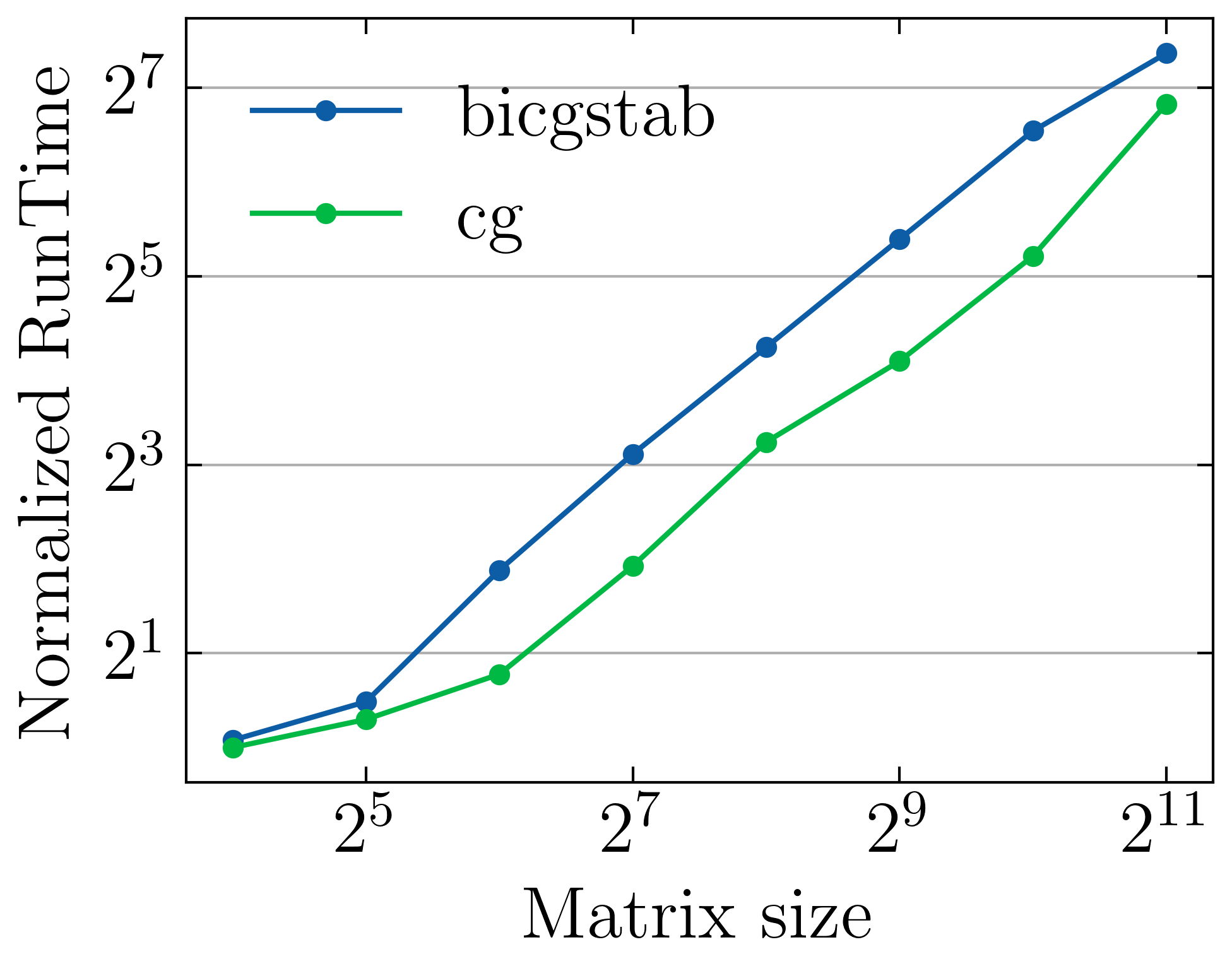}
  \caption{W.r.t matrix sizes,
           $2^{17}$ matrices}
  \label{fig:performance:scaling_matsize}
\end{subfigure}
\begin{subfigure}{\linewidth}
 \centering
  \includegraphics[width=0.65\linewidth]{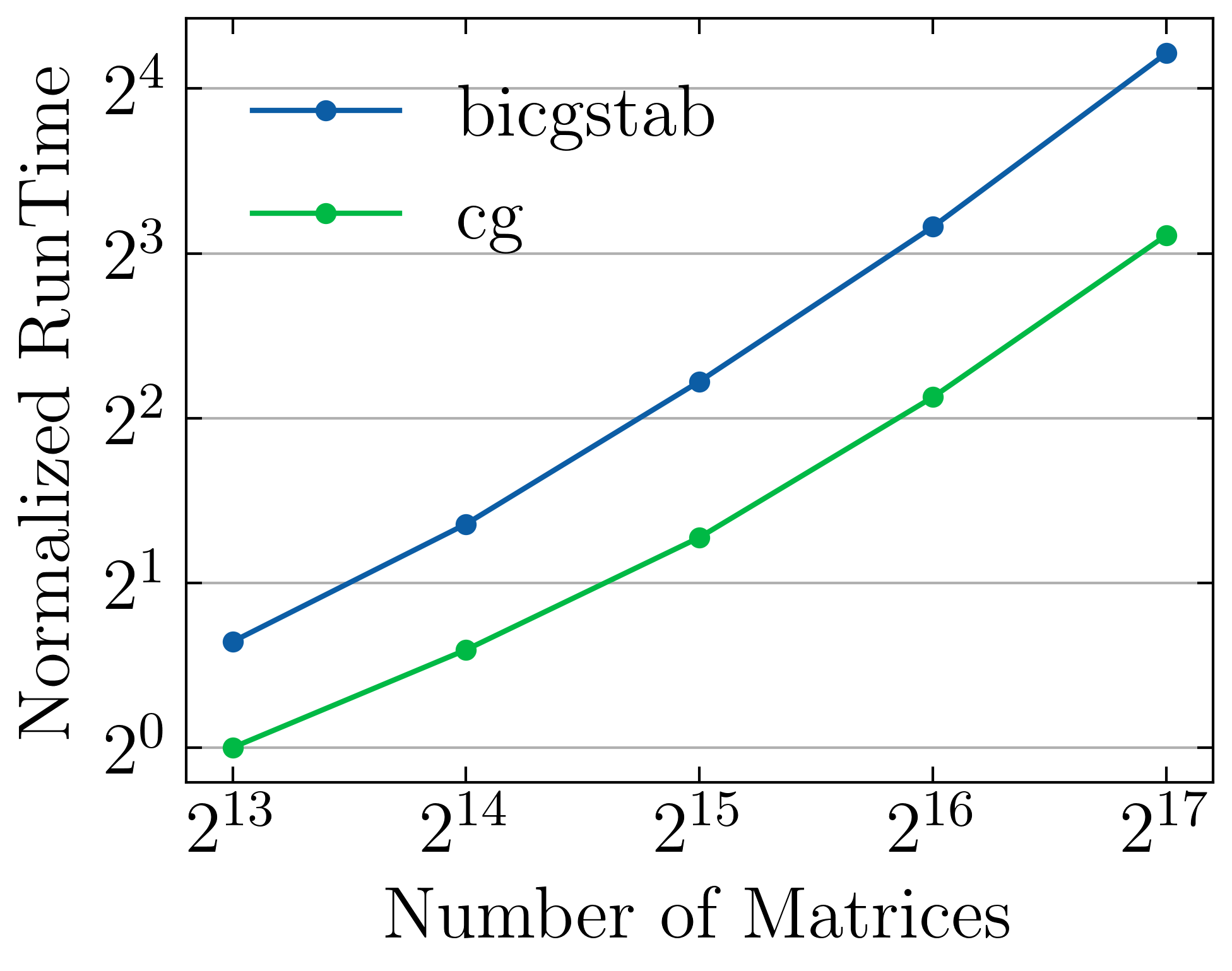}
  \caption{W.r.t number of matrices,
          the matrix size (64 x 64)}
  \label{fig:performance:scaling_nbatch}
\end{subfigure}
  \caption{Scaling of the SYCL batched solvers on 1 stack of the PVC GPU with
  respect to the problem sizes using the synthetic input. The runtimes of the
  solvers scale almost linearly with the problem size.}
  \label{fig:performance:scaling_problemsize}
\end{figure}

Using a standard 3-point stencil problem, we can generate a batch of symmetric,
positive definite (SPD) matrices that allows us to do scaling experiments in
both the matrix size and the batch size.
\Cref{fig:performance:scaling_problemsize} shows the scaling behaviour using 1
stack of the Intel GPU for both the \batchcg and the \batchbicgstab solver. In
\Cref{fig:performance:scaling_matsize}, we fix the batch size to \(2^{17}\) and
increase the matrix size (number of rows) for each of the batch items. As
expected, the overall runtime increases linearly with the matrix size. In
\Cref{fig:performance:scaling_nbatch}, we increase the number of items in the
batch from \(2^{13}\) to \(2^{17}\) for a individual problem size of \(64 \times
64\) and again observe a linear increase in the run-time. This means that we are
able to fully saturate the GPU , and additional linear systems need to wait for
the current systems to be completed. 
Overall, the great scalability of our batched solvers would allow them to
address different potential simulations which are problem-size dependent. 

\begin{figure}[t]
\centering
\begin{subfigure}{.775\linewidth}
 \centering
 \includegraphics[width=\linewidth]{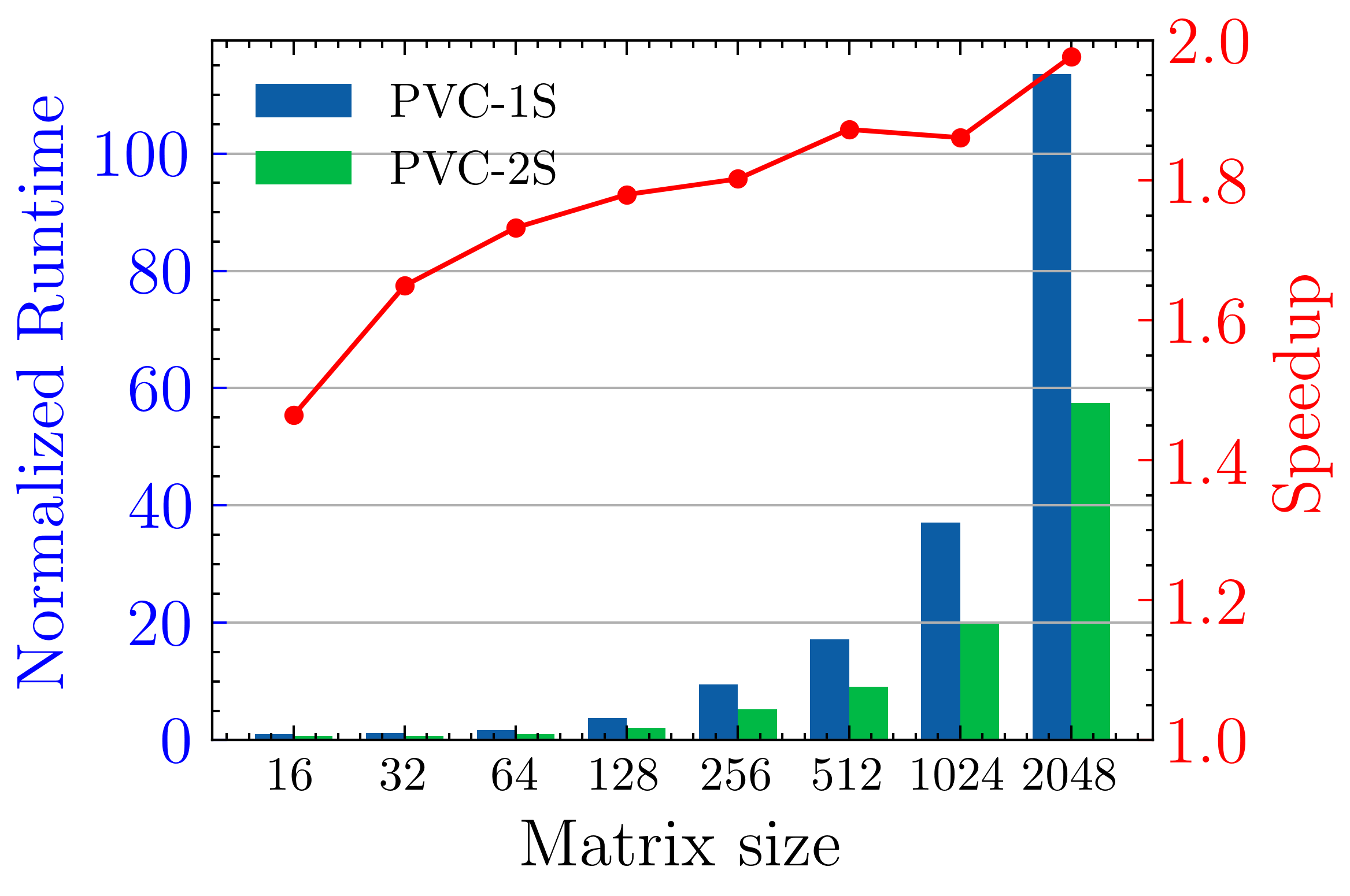}
 \caption{Batch CG}
 \label{fig:performance:scaling_cg}
\end{subfigure}
\begin{subfigure}{.775\linewidth}
 \centering
  \includegraphics[width=\linewidth]{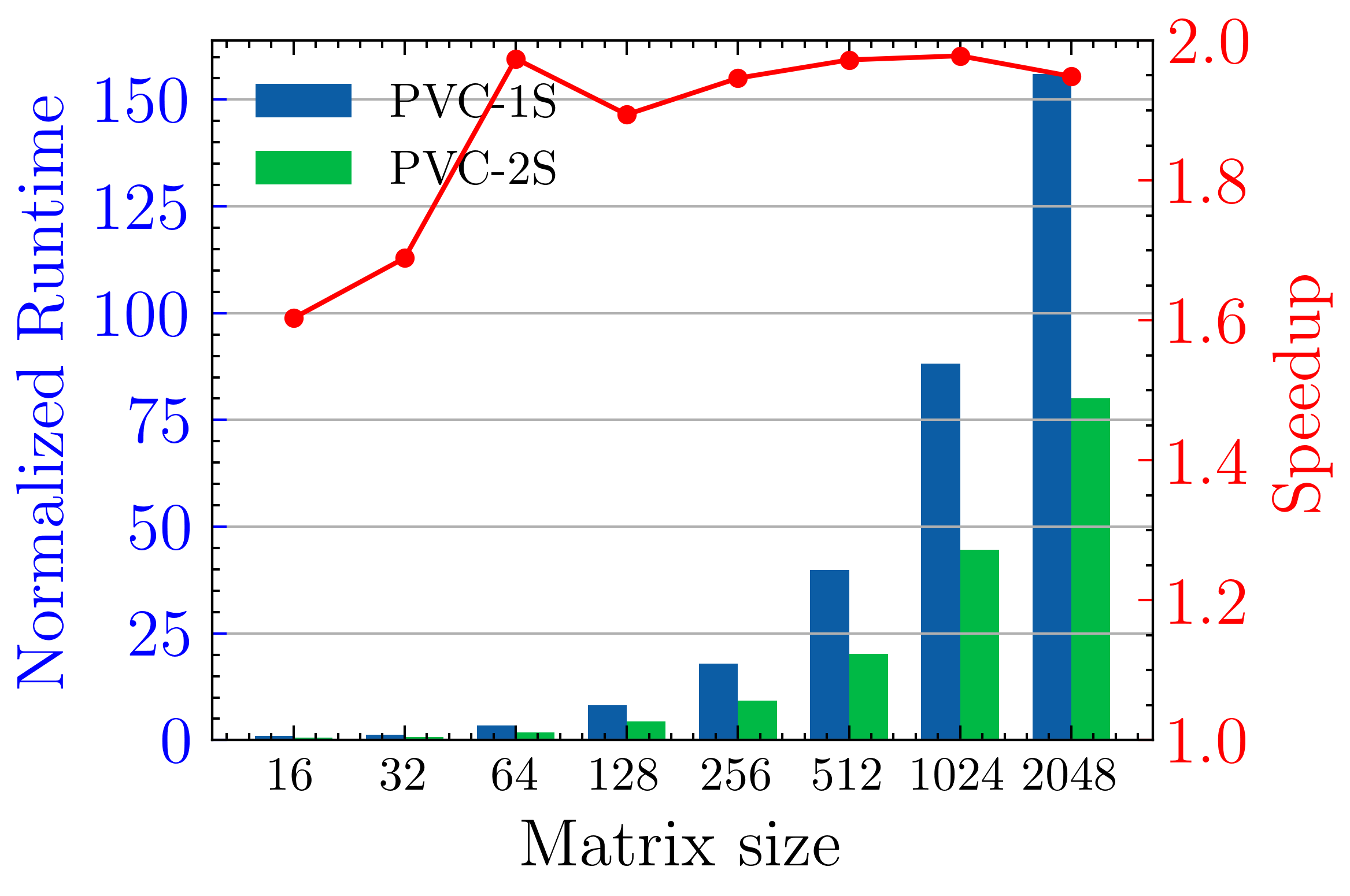}
 \caption{Batch BiCGSTAB}
 \label{fig:performance:scaling_bicg}
\end{subfigure}
\caption{Performance comparison of the SYCL batched solvers on 1- and 2-stacks
  of the PVC GPUs with respect to different matrix sizes. The synthetic input
  set is used with $2^{17}$ matrices. {\red For the speedup, the performance of
  1-stack is used as a baseline}. With implicit scaling on 2-stacks, the two
  solvers achieve between 1.5x -- 2.0x speedup, and the larger matrix size, the
  higher speedup.}
\label{fig:performance:scaling}
\end{figure}

The Intel GPU consists of two separate stacks that can be viewed as a single GPU
or as two separate GPUs, as previously mentioned. The batched solvers, being
embarrassing parallel, can take advantage of this feature via the implicit
scaling mode. It can be done automatically as the Intel GPU driver can split the
workloads, i.e. number of matrices, and schedule them on the two stacks without
explicit requests from the users.
This implicit scaling behaviour and its benefits are shown in \Cref{fig:performance:scaling} for both
\batchcg and \batchbicgstab solvers. We observe on average a 1.8x speedup for
\batchcg and 1.9x for the \batchbicgstab solver going from 1 stack to 2 stacks,
revealing that the batched solvers indeed provide the embarrassing parallelism
that can be harnessed by the GPUs. {\red The speedup, however, is lower than 2x due to
the NUMA effects, as the memory allocation is not perfectly split across the two
stacks with the \textit{implicit scaling} mode.}
Nevertheless, the reasonable achieved speedup suggests that we can easily scale
to multiple GPUs as distributing these batched matrices over the MPI ranks is
trivial and no additional communication is necessary.

\subsection{Performance evaluation with the real application inputs}

For benchmarking the performance of the batched solvers on the Intel GPUs,
against the batched solvers on state of the art NVIDIA GPUs, we use matrices
from the PeleLM+SUNDIALS application, described in
\Cref{tab:measurement:inputs}.  
Since these matrices are non-SPD, \batchcg can not be used to solve these systems,
thus only \batchbicgstab is evaluated in this section. The vendor-native
programming models are used: SYCL-based solvers for Intel PVC GPUs and
CUDA-based solvers for NVIDIA A100 and H100. The CUDA implementation of the batched
iterative solver can be found in our previous paper~\cite{aggarwalBatchedSparseIterative2021}.

\begin{figure}[t]
\centering
\begin{subfigure}{.5\linewidth}
 \centering
 \includegraphics[width=\linewidth]{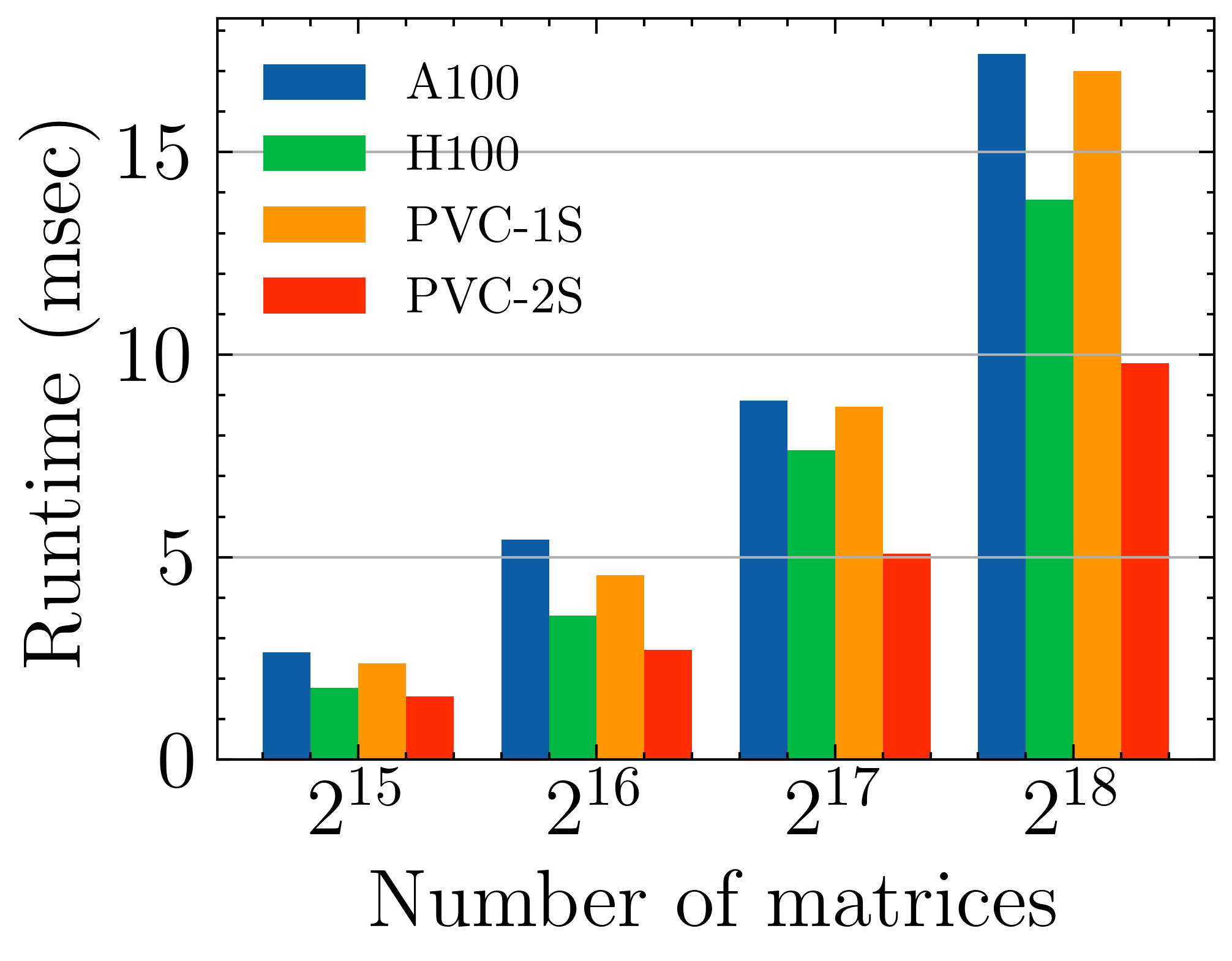}
  \caption{drm19 (Matrix size: 22 x 22)}
 \label{fig:performance:realinput_drm19}
\end{subfigure}%
\begin{subfigure}{.5\linewidth}
 \centering
 \includegraphics[width=\linewidth]{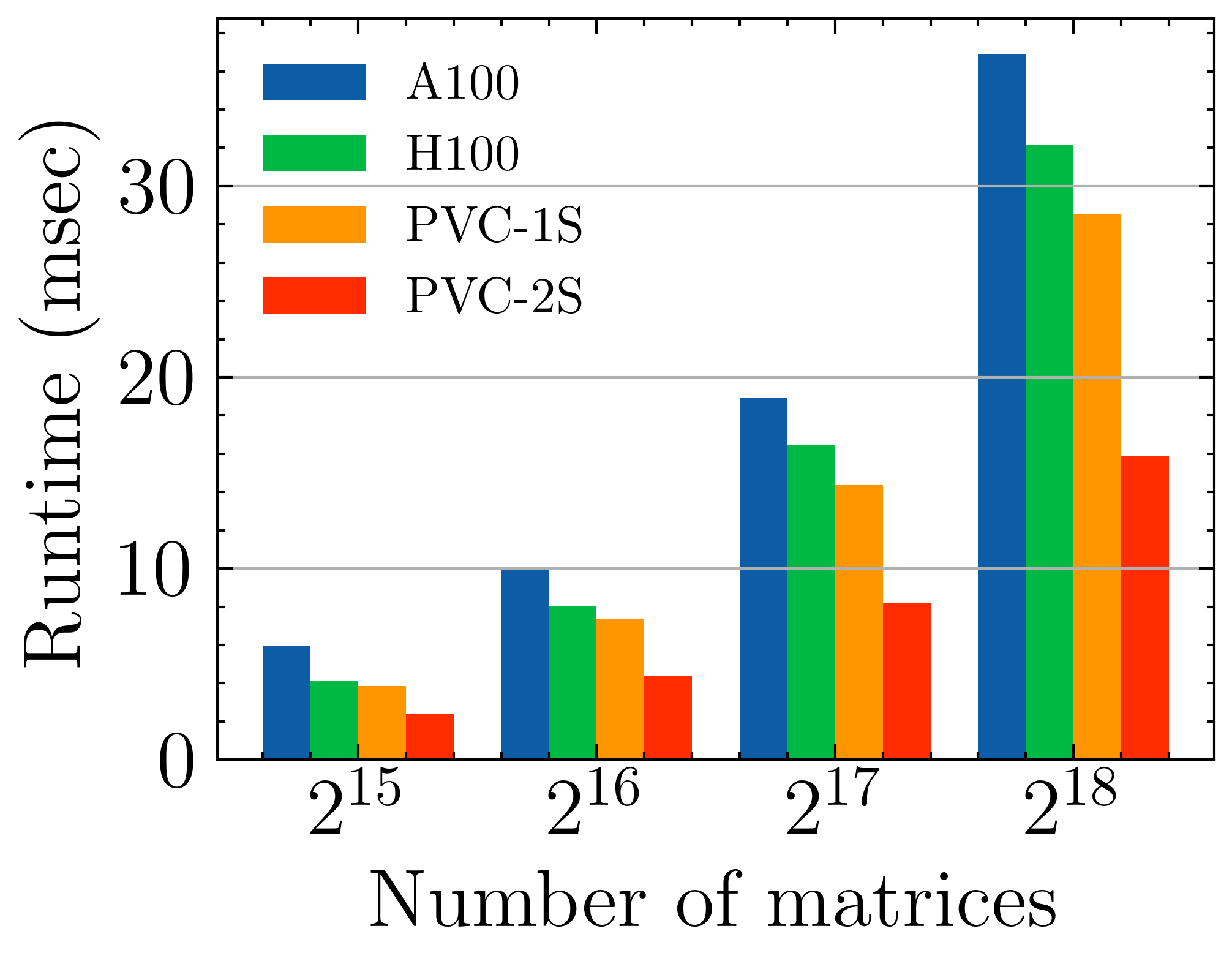}
  \caption{gri12 (Matrix size: 33 x 33)}
 \label{fig:performance:realinput_gri12}
\end{subfigure}
\begin{subfigure}{.5\linewidth}
 \centering
 \includegraphics[width=\linewidth]{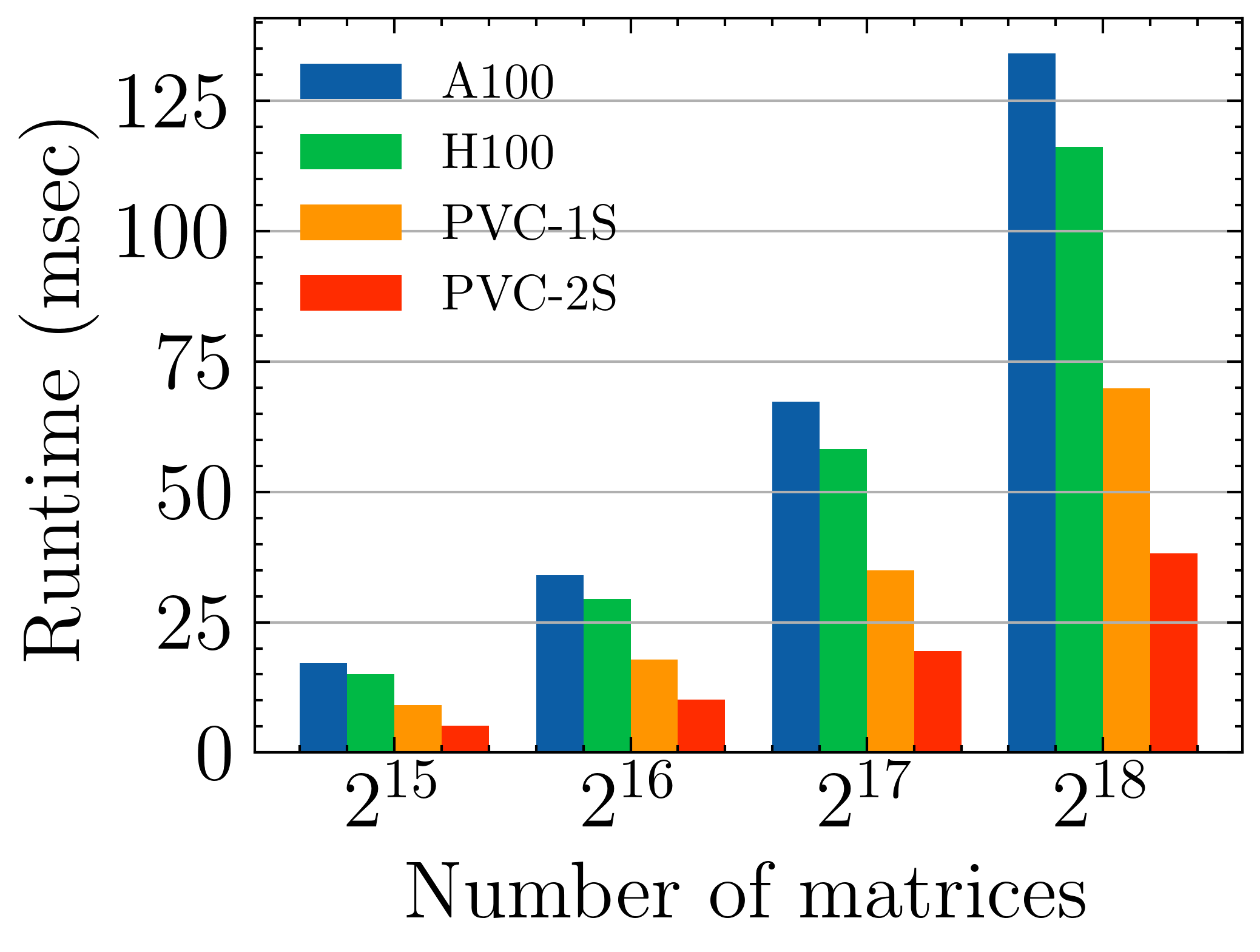}
  \caption{gri30 (Matrix size: 54 x 54)}
 \label{fig:performance:realinput_gri30}
\end{subfigure}%
\begin{subfigure}{.5\linewidth}
 \centering
 \includegraphics[width=\linewidth]{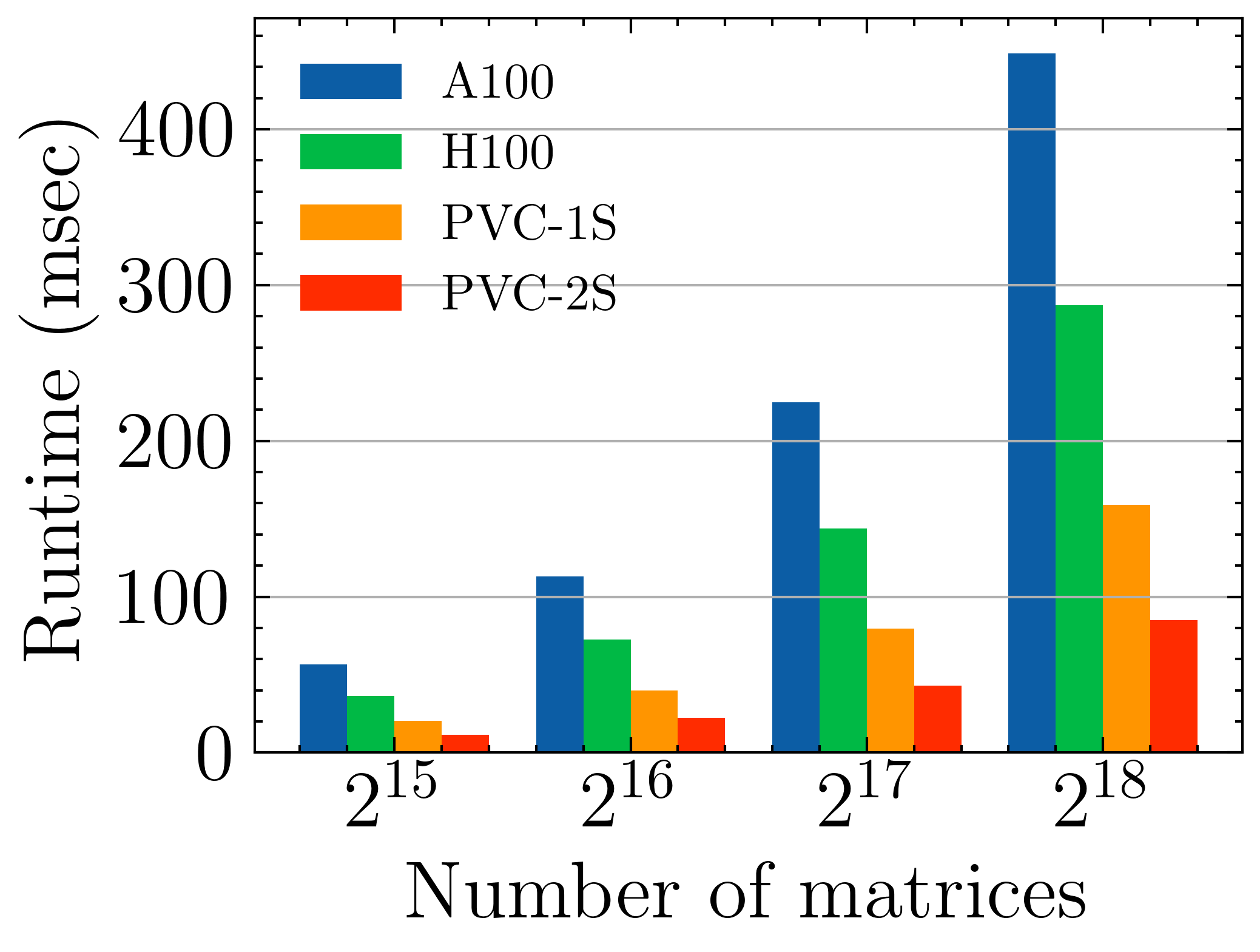}
  \caption{dodecane\_lu (Mat. size: 54 x 54)}
 \label{fig:performance:realinput_dodecane}
\end{subfigure}
\begin{subfigure}{.5\linewidth}
 \centering
 \includegraphics[width=\linewidth]{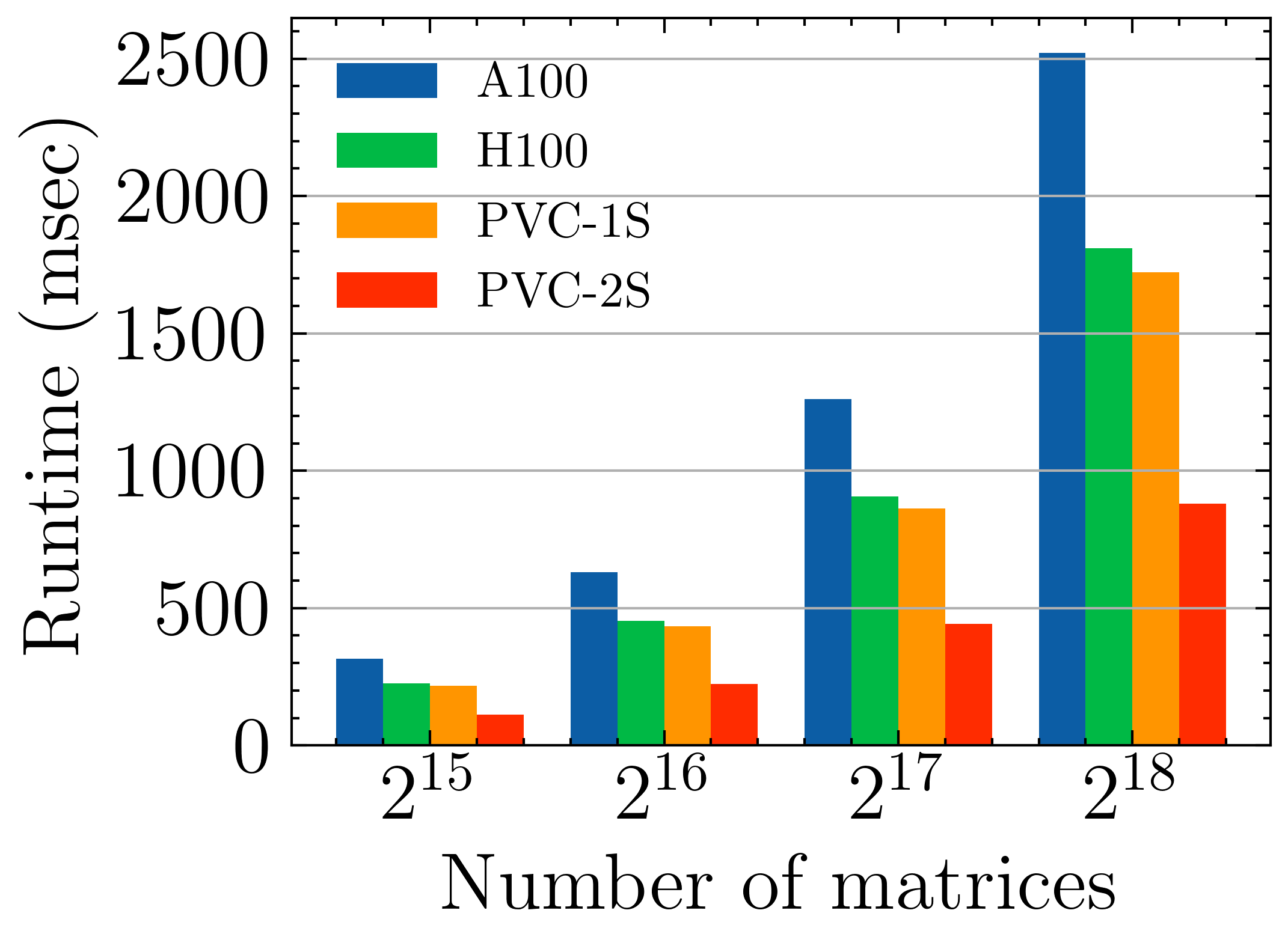}
  \caption{isooctane (Mat. size: 144 x 144)}
 \label{fig:performance:realinput_isooctane}
\end{subfigure}
\caption{Runtime of the two batched solvers on the three GPUs with different
  input data from the PeleLM simulations.
  Overall, the SYCL solvers on the PVC GPUs outperform the ones on the
  NVIDIA H100 for all input cases.}
\label{fig:performance:realinput_runtime}
\end{figure}

\Cref{fig:performance:realinput_runtime} presents the runtime comparison of the
solvers for solving the five test problems on NVIDIA A100, NVIDIA H100, and
Intel PVC GPUs. Overall, the batched solver on 2 stacks of the PVC
GPUs outperforms the ones on the A100 and H100 GPUs significantly for all
matrices and batch sizes.
In addition, the figure also demonstrates that the SYCL-based batched solvers
scale well on real application inputs in a similar fashion to the previous
experiment with the synthetic input matrices.

\begin{figure}[ht]
  \includegraphics[width=0.65\linewidth]{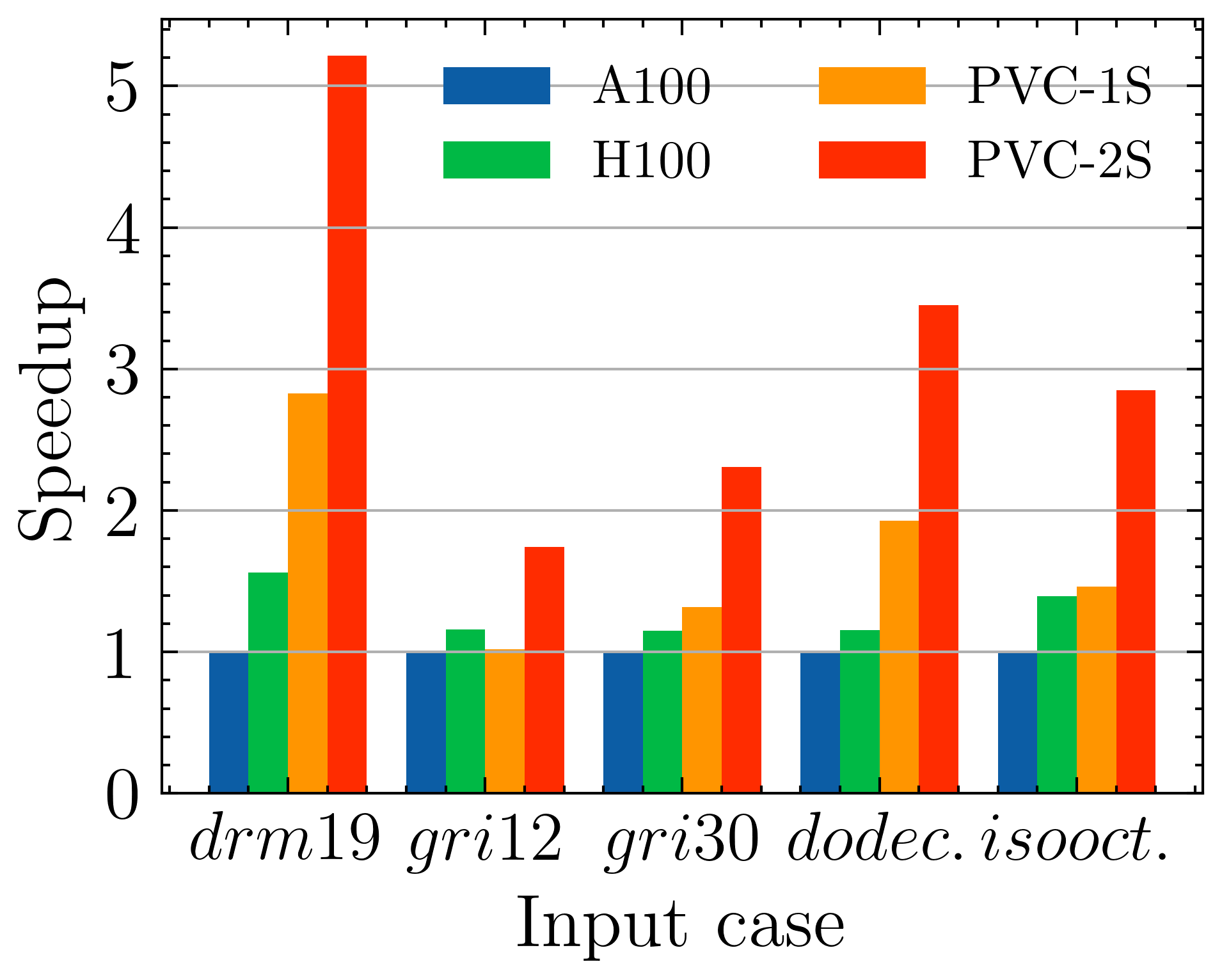}
  \caption{Normalized speedup comparison for different input cases with $2^{17}$
  matrices and the runtime of A100 is the baseline. The PVC GPU with 1- and
  2-stacks are 1.3 and 2.4 times faster then the H100, in average.}
  \label{fig:performance:realinput_speedup}
\end{figure}

\Cref{fig:performance:realinput_speedup} provides a direct performance
comparison of the batched solvers on the three GPUs.
Except for the \textit{gri12} case, all other other input cases shows notable
performance of the solvers on 1 stack of the PVC GPUs when compared with the NVIDIA
GPUs. 
In average, the PVC-1S is 1.7x and 1.3x
faster than the A100 and H100, respectively, across all input cases. 
Similarly, the PVC-2S outperforms the A100 and H100 by an average factor of 3.1
and 2.4, respectively.

\subsection{Roofline analysis}
\label{sec:analysis}
The performance of the SYCL-based batched solvers on the Intel GPUs
is also evaluated using the Intel Advisor Tool.

We analyse the \batchbicgstab solver for the \textit{dodecane\_lu} input case
with the batch size of $2^{17}$ matrices on 1 stack of the PVC GPU.  

The profiling results outline that the \gls{xve} Threading Occupancy is around
50\% with the \gls{xve} Array Active stays around 40\%. This means that the kernel
workloads do not fully occupy all available \gls{xve} on the \xe-cores.
This is expected, as in the solver kernel
implementation, we let each work-group use the maximum amount of
shared local memory available regardless of the work-group size.
With this strategy, even if the thread-group size is smaller than the number of
work-items a \xe-core can handle, there are not more thread-groups get scheduled
on the same \xe-core due to the limit on the available shared local memory. 
In other words, we trade the \gls{xve} occupancy for increased amount of
shared local memory usage in each work-group, which is more important
to achieve a good performance for the batched iterative solvers, as previously
discussed in \Cref{sec:impl:max_mem_usage}.

\begin{figure}[t]
  \includegraphics[width=\linewidth]{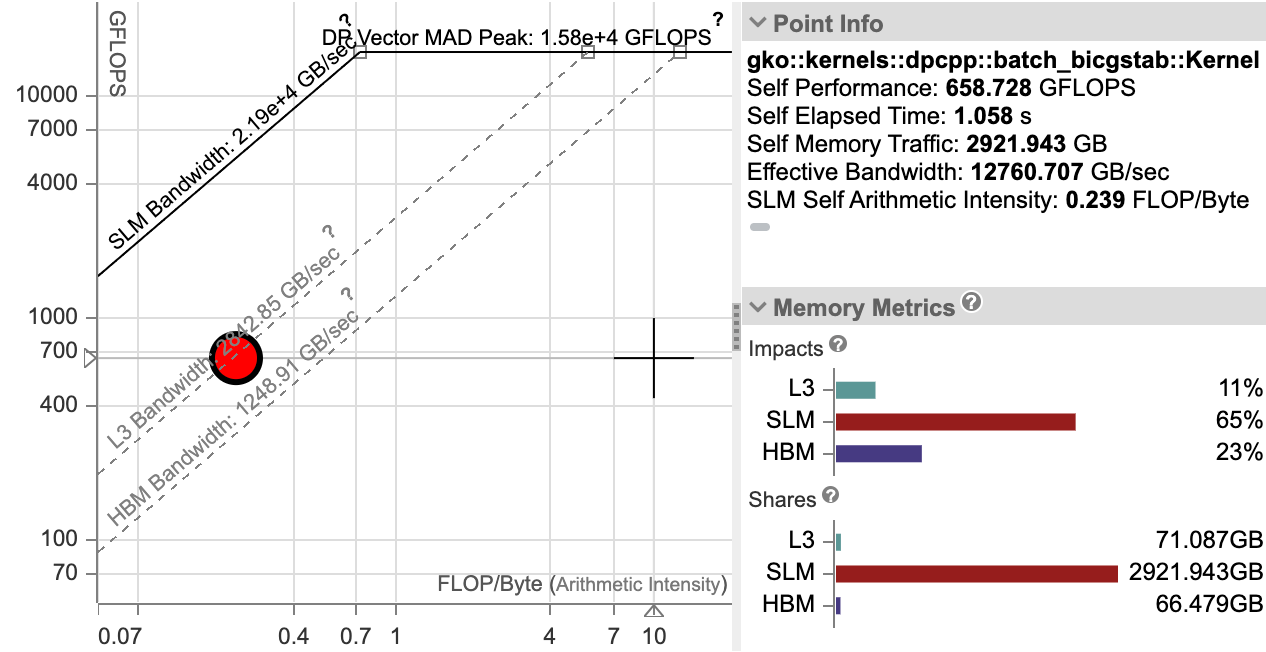}
  \caption{Roofline analysis and memory metrics of the \batchbicgstab for the \textit{dodecane\_lu}
  input case with $2^{17}$ matrices on 1-stack of the PVC GPU. The solver
  performance relies heavily on the Shared Local Memory and has not reach the
  SLM Bandwidth Bound.}
  \label{fig:analysis:roofline_dodecane_lu}
\end{figure}

\Cref{fig:analysis:roofline_dodecane_lu} presents the roofline performance
chart and the memory metrics of the solver. 
The time breakdown for the memory subsystem shows that 65\%  of the time spent
for memory transactions is spent on the SLM requests.
%
In addition, there are almost 3 TB of data passes through the SLM which is much
larger than the one passes through either the L3 or HBM.
This means that the performance of the solver mainly relies on the efficiency of the
SLM accesses which is expected, since the solver keeps all frequently accessed data in
the SLM for this input case. 
%
In addition, 11\% of memory accesses
are from the L3 (which is actually the L2 Cache on the GPU stack) implying that
the batch matrices and the right hand side (constant objects) are likely cached
into the this last level cache, which facilitates the accessing of these matrix
from the work-group.
In the roofline analysis, we observe that the performance of the solver lies on the L3
Bandwidth roof which is relatively good. However, the solver does not yet reach
the SLM Bandwidth roof. Further optimizations to improve SLM accesses,
for example identifying possible bank-conflicts and resolving them, will be part of our future work.


{\red
\subsection{Performance portability and productivity}
SYCL itself is a portable programming model and several applications use SYCL to
successfully support multiple hardware platforms, such as \cite{8945798,
10.1145/3529538.3529688}.
For the Ginkgo project, the existence of such a portable programming model
promises both performance portability and productivity for the library,
potentially giving us the opportunity to focus our work on algorithm
developments instead of supporting multiple backends.
Nonetheless, porting the Ginkgo SYCL backend to other platforms posed several
challenges, as discussed below.

Our first attempt was to port the Ginkgo SYCL backend, which includes both normal
routines and batched routines, onto the NVIDIA GPUs. 
We used the \textit{llvm} compiler with the SYCL extensions developed by Intel,
as it supports SYCL on the Intel GPUs, AMD GPUs, and NVIDIA
GPUs~\cite{oneAPIcompilers}.
Some components of the Ginkgo SYCL backend employ routines from the \gls{mkl}
and the \gls{dpl} which are partially supported on the NVIDIA
GPUs~\cite{MKLcompa}.  
We experienced an issue with linking against those Intel libraries while
compiling the Ginkgo SYCL backend for NVIDIA GPUs, an issue which is not
resolved at the time of this paper.
Since the implemented batched iterative solvers themselves do not use the routines from
neither \gls{mkl} or \gls{dpl}, ideally, we could compile them without linking
against these libraries. This can be done for portability evaluation
purposes, though other routines may not be fully functional.
However, even if this is done, the current state of the SYCL implementation
on CUDA platforms for the used llvm compiler does not fully support complex
floating-point functions~\cite{SYCLcomplex}, hence prohibiting us from
porting the routines onto the NVIDIA GPUs.

With the success of the A64FX-based Fugaku system at Riken, we have seen high
interest in sparse linear algebra libraries for ARM-based
systems~\cite{https://doi.org/10.1002/cpe.6512,9912708,
10.1145/3529538.3529688}. 
Additionally, SYCL compilers for ARM
processors have been developed~\cite{10.1145/3388333.3388658}. Therefore, we
attempted to port the Ginkgo SYCL backend for the A64FX processor. For this, we
used the OpenSYCL to compile our SYCL backend on the Ookami Cluster at Stony
Brook University via the ACCESS program~\cite{10.1145/3569951.3597559}.
We successfully verified the compiler and the setup environment with a simple standalone
kernel - \textit{vector-add}.
For compiling Ginkgo SYCL backend, we faced two main issues. First, \gls{mkl} and \gls{dpl} do not support
ARM processors. Linking against other ARM-supported math libraries requires
re-writing all API calls, as there is not yet a standard for sparse linear algebra
routines.
Second, while OpenSYCL has made significant progress in the implementation, it
has not achieved full SYCL conformance. The SYCL-based batched iterative
solvers, therefore, are not yet portable to the A64FX processors.


Nevertheless, Ginkgo supports multiple hardware architectures including CPUs,
NVIDIA GPUs, AMD GPUs, and Intel GPUs through OpenMP, CUDA, HIP, and SYCL
backends. Hence, Ginkgo itself is a performance portable library
from a user's perspective, as it provides users sparse solvers across multiple
platforms.
Even though the portability of the SYCL backend in general and the SYCL-based
batched iterative solvers in particular on other non-Intel GPUs platforms are
not achieved at the time of the paper, the developed SYCL-based solvers hold
promises for our future research endeavors in developing a portable backend
as well as improving the productivity of Ginkgo developers.
}

\section{Conclusion}
\label{sec:conclusion}

We have presented our successful work on porting the Ginkgo's batched iterative
solvers onto Intel GPUs with the SYCL programming model. 
%
The design of the batched iterative solvers is delineated and the different
optimization strategies are discussed.
The SYCL-based batched iterative solvers shows nearly linear scaling with
respect to the problem sizes and exhibit effective scaling on 2 stacks of the
PVC GPU using the implicit scaling mode.
The performance of the ported solvers on the Intel GPU Max 1550s surpasses the
one on NVIDIA H100 GPU by an average factor of 2.4 for the matrices from the PeleLM
application. 
Moreover, the solvers demonstrate exemplary the hardware resource utilization, as evidenced
by the analysis conducted using the Intel Advisor Tool.

\begin{appendices}
\section{Performance Appendix}
\label{appendix:performance}
\begin{table}[h!]
  \caption{\red Percentage speed-up when rounding the work-group size to a multiple of the sub-group size}
  \label{tab:appendix:roundup}
  \begin{center}
    \begin{tabular}[c]{lccccccc}
      \toprule
        Batch size &  32768 &  65536 &  131072
        & 262144 & Average\\
      \midrule
        \textit{drm19} & 0.8 & -0.2  & 0.6 & -0.6  & 1.3 \\
        \textit{gri12} & 50.9 & 54.8 & 50.8 & 52.1 & 48.5 \\
        \textit{gri30} & 2.9 & 2.6 & 3.2 & 3.1 & 3.2 \\
        \textit{dodecane\_lu} & 1.3 & 1.5 & 1.7   & 1.9 & 1.6 \\
        \textit{isooctane} & 0.0 & -0.1 & -0.0 & 0.1 & 0.1 \\
      \bottomrule
    \end{tabular}
  \end{center}
\end{table}

\section{Reproducibility Appendix}
\label{appendix:reproducibility}

In order to ensure reproducibility of results, we provide the code and elaborate on the settings and parameters used to produce these results.

\subsection*{\large Obtaining the source code}

The source code is open-source and available on the Ginkgo-Project Github
(\url{https://github.com/ginkgo-project/ginkgo.git}). The code used for this
paper is archived on Zenodo~\cite{nguyenReproducibilityArtifactGinkgo2023}.

\subsection*{\large Building and installing Ginkgo}

To build \gko with SYCL, the following components are necessary:
\begin{enumerate}
\item The CMake build platform, CMake-3.26.3 was used in this paper.
\item Intel oneAPI Toolkit installation, Intel oneAPI 2023.05.15.006 was used in
  this paper.
\end{enumerate}

The Ginkgo library and the batched functionality use the same canonical CMake setup as elaborated in the Ginkgo documentation (\url{https://ginkgo-project.github.io/ginkgo/doc/develop/install_ginkgo.html}).

\subsection*{\large Benchmarking}

The performance results from this paper can be reproduced following
these steps:

\begin{enumerate}
  \item Building \gko with SYCL backend:
    \begin{itemize}
      \item Make a build directory: \texttt{\$ mkdir build \&\& cd build}
      \item Configure \gko with the SYCL backend : \\
        \texttt{\$ cmake -DCMAKE\_CXX\_COMPILER=icpx \newline -DGINKGO\_BUILD\_DPCPP=on ..}
      \item Compile \gko: 
        \texttt{\$ make}
    \end{itemize}
  \item Building \gko with CUDA backend: follow the reproducibility appendix in
    \cite{aggarwalBatchedSparseIterative2021}.
  \item Performance tests:
    \begin{itemize}
      \item The benchmarks with the synthetic 3pt stencil input can be found in
        the directory: \newline
        \texttt{./ginkgo/examples/batched-solver}.
      \item The benchmarks with input matrices from the PeleLM application can
        be found in the directory: \newline
        \texttt{ginkgo/examples/batched-solver-from-files}.
      \item The benchmarking scripts for both input classes are provided in
        \texttt{run-test-dpcpp.sh} and \texttt{run-test-cuda.sh}.
    \end{itemize}
\end{enumerate}

\subsection*{\large Our setup}

The performance of the solvers on the Intel GPUs were measured on the Sunspot -
the testbed system for the Aurora supercomputer - deployed by the Argonne
Leadership Computing Facility at Argonne National Laboratory, US. 
Each node consists of 2x Intel Xeon CPU Max Series (Sapphire Rapids)
and 6x Intel Data Center GPU Max Series (PVC).
Compilers and related libraries are from the Intel oneAPI 2023.05.15.006 Toolkit. 

The performance of the solvers on the NVIDIA GPUs were obtained from our
in-house testing nodes at the University of Tennessee, Knoxville, as follow:
\begin{itemize}
  \item 1 node: 2x Intel Xeon Silver 4309Y CPU and 1x NVIDIA H100 PCIe Gen5
    80GB.
  \item 1 node: 2x AMD EPYC 7742 CPU and 8x NVIDIA A100 SXM4 80GB .
\end{itemize}
GCC-11.3.1 was used as the host compiler and CUDA Toolkit 11.8.0 was used as the device
compiler on both nodes.

\end{appendices}

\section{Acknowledgments}
This research was supported by the Exascale Computing Project (17-SC-20-SC), a
collaborative effort of the U.S. Department of Energy Office of Science and the
National Nuclear Security Administration. 
In addition, this work used the Sunspot testbed under the CLOVER, XSDK projects
at the Argonne Leadership Computing Facility, which is a DOE Office of Science
User Facility, and the Ookami cluster at Stony Brook university through
allocation CIS230121 from the Advanced Cyberinfrastructure Coordination
Ecosystem: Services \& Support (ACCESS) program, which is supported by National
Science Foundation grants \#2138259, \#2138286, \#2138307, \#2137603, and \#2138296.

Special thanks to Thomas Applencourt and Abhishek Bagusetty for their valuable
supports on working with SYCL and the Intel GPUs, Jens Domke for insights and
approaches on porting code to A64FX, and John Pennycook for shepherding the
paper.

\balance
\bibliographystyle{ACM-Reference-Format}

\begin{thebibliography}{36}


\ifx \showCODEN    \undefined \def \showCODEN     #1{\unskip}     \fi
\ifx \showDOI      \undefined \def \showDOI       #1{#1}\fi
\ifx \showISBNx    \undefined \def \showISBNx     #1{\unskip}     \fi
\ifx \showISBNxiii \undefined \def \showISBNxiii  #1{\unskip}     \fi
\ifx \showISSN     \undefined \def \showISSN      #1{\unskip}     \fi
\ifx \showLCCN     \undefined \def \showLCCN      #1{\unskip}     \fi
\ifx \shownote     \undefined \def \shownote      #1{#1}          \fi
\ifx \showarticletitle \undefined \def \showarticletitle #1{#1}   \fi
\ifx \showURL      \undefined \def \showURL       {\relax}        \fi
\providecommand\bibfield[2]{#2}
\providecommand\bibinfo[2]{#2}
\providecommand\natexlab[1]{#1}
\providecommand\showeprint[2][]{arXiv:#2}

\bibitem[MKL(2020)]%
        {MKLcompa}
 \bibinfo{year}{2020}\natexlab{}.
\newblock \bibinfo{title}{[CUDA] MKL compatibility}.
\newblock
  \bibinfo{howpublished}{\url{https://github.com/intel/llvm/issues/1548}}.
\newblock


\bibitem[SYC(2023)]%
        {SYCLcomplex}
 \bibinfo{year}{2023}\natexlab{}.
\newblock \bibinfo{title}{[SYCL][CUDA] ptxas fatal: complex floating-point
  functions}.
\newblock
  \bibinfo{howpublished}{\url{https://github.com/intel/llvm/issues/8281}}.
\newblock


\bibitem[Abdelfattah et~al\mbox{.}(2021)]%
        {10.1145/3431921}
\bibfield{author}{\bibinfo{person}{Ahmad Abdelfattah}, \bibinfo{person}{Timothy
  Costa}, \bibinfo{person}{Jack Dongarra}, \bibinfo{person}{Mark Gates},
  \bibinfo{person}{Azzam Haidar}, \bibinfo{person}{Sven Hammarling},
  \bibinfo{person}{Nicholas~J. Higham}, \bibinfo{person}{Jakub Kurzak},
  \bibinfo{person}{Piotr Luszczek}, \bibinfo{person}{Stanimire Tomov}, {and}
  \bibinfo{person}{Mawussi Zounon}.} \bibinfo{year}{2021}\natexlab{}.
\newblock \showarticletitle{A Set of Batched Basic Linear Algebra Subprograms
  and {LAPACK} Routines}.
\newblock \bibinfo{journal}{\emph{ACM Trans. Math. Softw.}}
  \bibinfo{volume}{47}, \bibinfo{number}{3}, Article \bibinfo{articleno}{21}
  (\bibinfo{date}{June} \bibinfo{year}{2021}), \bibinfo{numpages}{23}~pages.
\newblock
\showISSN{0098-3500}
\urldef\tempurl%
\url{https://doi.org/10.1145/3431921}
\showDOI{\tempurl}


\bibitem[Abdelfattah et~al\mbox{.}(2016)]%
        {DBLP:conf/supercomputer/AbdelfattahHTD16}
\bibfield{author}{\bibinfo{person}{Ahmad Abdelfattah}, \bibinfo{person}{Azzam
  Haidar}, \bibinfo{person}{Stanimire Tomov}, {and} \bibinfo{person}{Jack~J.
  Dongarra}.} \bibinfo{year}{2016}\natexlab{}.
\newblock \showarticletitle{Performance, Design, and Autotuning of Batched
  {GEMM} for GPUs}. In \bibinfo{booktitle}{\emph{High Performance Computing -
  31st International Conference, {ISC} High Performance 2016, Frankfurt,
  Germany, June 19-23, 2016, Proceedings}} \emph{(\bibinfo{series}{Lecture
  Notes in Computer Science}, Vol.~\bibinfo{volume}{9697})},
  \bibfield{editor}{\bibinfo{person}{Julian~M. Kunkel}, \bibinfo{person}{Pavan
  Balaji}, {and} \bibinfo{person}{Jack~J. Dongarra}} (Eds.).
  \bibinfo{publisher}{Springer}, \bibinfo{pages}{21--38}.
\newblock
\urldef\tempurl%
\url{https://doi.org/10.1007/978-3-319-41321-1\_2}
\showDOI{\tempurl}


\bibitem[Aggarwal et~al\mbox{.}(2021)]%
        {aggarwalBatchedSparseIterative2021}
\bibfield{author}{\bibinfo{person}{Isha Aggarwal}, \bibinfo{person}{Aditya
  Kashi}, \bibinfo{person}{Pratik Nayak}, \bibinfo{person}{Cody~J. Balos},
  \bibinfo{person}{Carol~S. Woodward}, {and} \bibinfo{person}{Hartwig Anzt}.}
  \bibinfo{year}{2021}\natexlab{}.
\newblock \showarticletitle{Batched {{Sparse Iterative Solvers}} for
  {{Computational Chemistry Simulations}} on {{GPUs}}}. In
  \bibinfo{booktitle}{\emph{2021 12th {{Workshop}} on {{Latest Advances}} in
  {{Scalable Algorithms}} for {{Large-Scale Systems}} ({{ScalA}})}}.
  \bibinfo{pages}{35--43}.
\newblock
\urldef\tempurl%
\url{https://doi.org/10/gn3xcg}
\showDOI{\tempurl}


\bibitem[Alappat et~al\mbox{.}(2022)]%
        {https://doi.org/10.1002/cpe.6512}
\bibfield{author}{\bibinfo{person}{Christie Alappat}, \bibinfo{person}{Nils
  Meyer}, \bibinfo{person}{Jan Laukemann}, \bibinfo{person}{Thomas Gruber},
  \bibinfo{person}{Georg Hager}, \bibinfo{person}{Gerhard Wellein}, {and}
  \bibinfo{person}{Tilo Wettig}.} \bibinfo{year}{2022}\natexlab{}.
\newblock \showarticletitle{Execution-Cache-Memory modeling and performance
  tuning of sparse matrix-vector multiplication and Lattice quantum
  chromodynamics on A64FX}.
\newblock \bibinfo{journal}{\emph{Concurrency and Computation: Practice and
  Experience}} \bibinfo{volume}{34}, \bibinfo{number}{20}
  (\bibinfo{year}{2022}), \bibinfo{pages}{e6512}.
\newblock
\urldef\tempurl%
\url{https://doi.org/10.1002/cpe.6512}
\showDOI{\tempurl}
\showeprint{https://onlinelibrary.wiley.com/doi/pdf/10.1002/cpe.6512}


\bibitem[Alpay and Heuveline(2020)]%
        {10.1145/3388333.3388658}
\bibfield{author}{\bibinfo{person}{Aksel Alpay} {and} \bibinfo{person}{Vincent
  Heuveline}.} \bibinfo{year}{2020}\natexlab{}.
\newblock \showarticletitle{SYCL beyond OpenCL: The Architecture, Current State
  and Future Direction of HipSYCL}. In \bibinfo{booktitle}{\emph{Proceedings of
  the International Workshop on OpenCL}} (Munich, Germany)
  \emph{(\bibinfo{series}{IWOCL '20})}. \bibinfo{publisher}{Association for
  Computing Machinery}, \bibinfo{address}{New York, NY, USA}, Article
  \bibinfo{articleno}{8}, \bibinfo{numpages}{1}~pages.
\newblock
\showISBNx{9781450375313}
\urldef\tempurl%
\url{https://doi.org/10.1145/3388333.3388658}
\showDOI{\tempurl}


\bibitem[Anzt et~al\mbox{.}(2016)]%
        {Anzt2016}
\bibfield{author}{\bibinfo{person}{Hartwig Anzt}, \bibinfo{person}{Edmond
  Chow}, \bibinfo{person}{Thomas Huckle}, {and} \bibinfo{person}{Jack
  Dongarra}.} \bibinfo{year}{2016}\natexlab{}.
\newblock \showarticletitle{Batched Generation of Incomplete Sparse Approximate
  Inverses on GPUs}. In \bibinfo{booktitle}{\emph{2016 7th Workshop on Latest
  Advances in Scalable Algorithms for Large-Scale Systems (ScalA)}}.
  \bibinfo{pages}{49--56}.
\newblock
\urldef\tempurl%
\url{https://doi.org/10.1109/ScalA.2016.011}
\showDOI{\tempurl}


\bibitem[Anzt et~al\mbox{.}(2020a)]%
        {ginkgo-joss}
\bibfield{author}{\bibinfo{person}{Hartwig Anzt}, \bibinfo{person}{Terry
  Cojean}, \bibinfo{person}{Yen-Chen Chen}, \bibinfo{person}{Goran Flegar},
  \bibinfo{person}{Fritz G{\"o}bel}, \bibinfo{person}{Thomas Gr{\"u}tzmacher},
  \bibinfo{person}{Pratik Nayak}, \bibinfo{person}{Tobias Ribizel}, {and}
  \bibinfo{person}{Yu-Hsiang Tsai}.} \bibinfo{year}{2020}\natexlab{a}.
\newblock \showarticletitle{Ginkgo: A High Performance Numerical Linear Algebra
  Library}.
\newblock \bibinfo{journal}{\emph{Journal of Open Source Software}}
  (\bibinfo{date}{Aug.} \bibinfo{year}{2020}).
\newblock
\urldef\tempurl%
\url{https://doi.org/10.21105/joss.02260}
\showDOI{\tempurl}


\bibitem[Anzt et~al\mbox{.}(2020b)]%
        {ginkgo-arxiv}
\bibfield{author}{\bibinfo{person}{Hartwig Anzt}, \bibinfo{person}{Terry
  Cojean}, \bibinfo{person}{Goran Flegar}, \bibinfo{person}{Fritz G{\"o}bel},
  \bibinfo{person}{Thomas Gr{\"u}tzmacher}, \bibinfo{person}{Pratik Nayak},
  \bibinfo{person}{Tobias Ribizel}, \bibinfo{person}{Yuhsiang~Mike Tsai}, {and}
  \bibinfo{person}{Enrique~S. {Quintana-Ort{\'i}}}.}
  \bibinfo{year}{2020}\natexlab{b}.
\newblock \showarticletitle{Ginkgo: {{A Modern Linear Operator Algebra
  Framework}} for {{High Performance Computing}}}.
\newblock \bibinfo{journal}{\emph{arXiv:2006.16852 [cs]}} (\bibinfo{date}{July}
  \bibinfo{year}{2020}).
\newblock
\showeprint[arxiv]{2006.16852}~[cs]


\bibitem[Anzt et~al\mbox{.}(2017a)]%
        {10.1145/3148226.3148230}
\bibfield{author}{\bibinfo{person}{Hartwig Anzt}, \bibinfo{person}{Gary
  Collins}, \bibinfo{person}{Jack Dongarra}, \bibinfo{person}{Goran Flegar},
  {and} \bibinfo{person}{Enrique~S. Quintana-Ort\'{\i}}.}
  \bibinfo{year}{2017}\natexlab{a}.
\newblock \showarticletitle{Flexible Batched Sparse Matrix-Vector Product on
  {GPUs}}. In \bibinfo{booktitle}{\emph{Proceedings of the 8th Workshop on
  Latest Advances in Scalable Algorithms for Large-Scale Systems}} (Denver,
  Colorado) \emph{(\bibinfo{series}{ScalA '17})}.
  \bibinfo{publisher}{Association for Computing Machinery},
  \bibinfo{address}{New York, NY, USA}, Article \bibinfo{articleno}{3},
  \bibinfo{numpages}{8}~pages.
\newblock
\showISBNx{9781450351256}
\urldef\tempurl%
\url{https://doi.org/10.1145/3148226.3148230}
\showDOI{\tempurl}


\bibitem[Anzt et~al\mbox{.}(2017b)]%
        {batchedlu}
\bibfield{author}{\bibinfo{person}{Hartwig Anzt}, \bibinfo{person}{Jack
  Dongarra}, \bibinfo{person}{Goran Flegar}, {and} \bibinfo{person}{Enrique~S.
  Quintana-Ortí}.} \bibinfo{year}{2017}\natexlab{b}.
\newblock \showarticletitle{Variable-Size Batched {LU} for Small Matrices and
  Its Integration into Block-{J}acobi Preconditioning}. In
  \bibinfo{booktitle}{\emph{2017 46th International Conference on Parallel
  Processing (ICPP)}}. \bibinfo{pages}{91--100}.
\newblock
\urldef\tempurl%
\url{https://doi.org/10.1109/ICPP.2017.18}
\showDOI{\tempurl}


\bibitem[Applencourt et~al\mbox{.}(2023)]%
        {10.1145/3585341.3585343}
\bibfield{author}{\bibinfo{person}{Thomas Applencourt}, \bibinfo{person}{Brice
  Videau}, \bibinfo{person}{Jefferson Le~Quellec}, \bibinfo{person}{Amanda
  Dufek}, \bibinfo{person}{Kevin Harms}, \bibinfo{person}{Nevin Liber},
  \bibinfo{person}{Bryce Allen}, {and} \bibinfo{person}{Aiden Belton-Schure}.}
  \bibinfo{year}{2023}\natexlab{}.
\newblock \showarticletitle{Standardizing Complex Numbers in SYCL}. In
  \bibinfo{booktitle}{\emph{Proceedings of the 2023 International Workshop on
  OpenCL}} (Cambridge, United Kingdom) \emph{(\bibinfo{series}{IWOCL '23})}.
  \bibinfo{publisher}{Association for Computing Machinery},
  \bibinfo{address}{New York, NY, USA}, Article \bibinfo{articleno}{2},
  \bibinfo{numpages}{6}~pages.
\newblock
\showISBNx{9798400707452}
\urldef\tempurl%
\url{https://doi.org/10.1145/3585341.3585343}
\showDOI{\tempurl}


\bibitem[Boerner et~al\mbox{.}(2023)]%
        {10.1145/3569951.3597559}
\bibfield{author}{\bibinfo{person}{Timothy~J. Boerner},
  \bibinfo{person}{Stephen Deems}, \bibinfo{person}{Thomas~R. Furlani},
  \bibinfo{person}{Shelley~L. Knuth}, {and} \bibinfo{person}{John Towns}.}
  \bibinfo{year}{2023}\natexlab{}.
\newblock \showarticletitle{ACCESS: Advancing Innovation: NSF’s Advanced
  Cyberinfrastructure Coordination Ecosystem: Services \& Support}. In
  \bibinfo{booktitle}{\emph{Practice and Experience in Advanced Research
  Computing}} (Portland, OR, USA) \emph{(\bibinfo{series}{PEARC '23})}.
  \bibinfo{publisher}{Association for Computing Machinery},
  \bibinfo{address}{New York, NY, USA}, \bibinfo{pages}{173–176}.
\newblock
\showISBNx{9781450399852}
\urldef\tempurl%
\url{https://doi.org/10.1145/3569951.3597559}
\showDOI{\tempurl}


\bibitem[Boukaram et~al\mbox{.}(2018)]%
        {10.1016/j.parco.2017.09.001}
\bibfield{author}{\bibinfo{person}{Wajih~Halim Boukaram},
  \bibinfo{person}{George Turkiyyah}, \bibinfo{person}{Hatem Ltaief}, {and}
  \bibinfo{person}{David~E. Keyes}.} \bibinfo{year}{2018}\natexlab{}.
\newblock \showarticletitle{Batched {QR} and {SVD} Algorithms on {GPUs} with
  Applications in Hierarchical Matrix Compression}.
\newblock \bibinfo{journal}{\emph{Parallel Comput.}} \bibinfo{volume}{74},
  \bibinfo{number}{C} (\bibinfo{date}{May} \bibinfo{year}{2018}),
  \bibinfo{pages}{19–33}.
\newblock
\showISSN{0167-8191}
\urldef\tempurl%
\url{https://doi.org/10.1016/j.parco.2017.09.001}
\showDOI{\tempurl}


\bibitem[Carroll et~al\mbox{.}(2021a)]%
        {carrollBatchedGPUMethodology2021a}
\bibfield{author}{\bibinfo{person}{Enda Carroll}, \bibinfo{person}{Andrew
  Gloster}, \bibinfo{person}{Miguel~D. Bustamante}, {and}
  \bibinfo{person}{Lennon~{\'O}' N{\'a}raigh}.}
  \bibinfo{year}{2021}\natexlab{a}.
\newblock \bibinfo{title}{A {{Batched GPU Methodology}} for {{Numerical
  Solutions}} of {{Partial Differential Equations}}}.
\newblock
\newblock
\urldef\tempurl%
\url{https://doi.org/10.48550/arXiv.2107.05395}
\showDOI{\tempurl}
\showeprint[arxiv]{2107.05395}~[physics]


\bibitem[Carroll et~al\mbox{.}(2021b)]%
        {carroll2021batched}
\bibfield{author}{\bibinfo{person}{Enda Carroll}, \bibinfo{person}{Andrew
  Gloster}, \bibinfo{person}{Miguel~D. Bustamante}, {and}
  \bibinfo{person}{Lennon~Ó' Náraigh}.} \bibinfo{year}{2021}\natexlab{b}.
\newblock \showarticletitle{A Batched {GPU} Methodology for Numerical Solutions
  of Partial Differential Equations}.
\newblock \bibinfo{journal}{\emph{arXiv}}  \bibinfo{volume}{2107.05395}
  (\bibinfo{year}{2021}).
\newblock
\showeprint[arxiv]{2107.05395}~[physics.comp-ph]


\bibitem[Crisci et~al\mbox{.}(2022)]%
        {10.1145/3529538.3529688}
\bibfield{author}{\bibinfo{person}{Luigi Crisci}, \bibinfo{person}{Majid
  Salimi~Beni}, \bibinfo{person}{Biagio Cosenza}, \bibinfo{person}{Nicol\`{o}
  Scipione}, \bibinfo{person}{Davide Gadioli}, \bibinfo{person}{Emanuele
  Vitali}, \bibinfo{person}{Gianluca Palermo}, {and} \bibinfo{person}{Andrea
  Beccari}.} \bibinfo{year}{2022}\natexlab{}.
\newblock \showarticletitle{Towards a Portable Drug Discovery Pipeline with
  SYCL 2020}. In \bibinfo{booktitle}{\emph{International Workshop on OpenCL}}
  (Bristol, United Kingdom, United Kingdom)
  \emph{(\bibinfo{series}{IWOCL'22})}. \bibinfo{publisher}{Association for
  Computing Machinery}, \bibinfo{address}{New York, NY, USA}, Article
  \bibinfo{articleno}{5}, \bibinfo{numpages}{2}~pages.
\newblock
\showISBNx{9781450396585}
\urldef\tempurl%
\url{https://doi.org/10.1145/3529538.3529688}
\showDOI{\tempurl}


\bibitem[Deakin et~al\mbox{.}(2020)]%
        {9309052}
\bibfield{author}{\bibinfo{person}{Tom Deakin}, \bibinfo{person}{Andrei
  Poenaru}, \bibinfo{person}{Tom Lin}, {and} \bibinfo{person}{Simon
  McIntosh-Smith}.} \bibinfo{year}{2020}\natexlab{}.
\newblock \showarticletitle{Tracking Performance Portability on the Yellow
  Brick Road to Exascale}. In \bibinfo{booktitle}{\emph{2020 IEEE/ACM
  International Workshop on Performance, Portability and Productivity in HPC
  (P3HPC)}}. \bibinfo{pages}{1--13}.
\newblock
\urldef\tempurl%
\url{https://doi.org/10.1109/P3HPC51967.2020.00006}
\showDOI{\tempurl}


\bibitem[Hindmarsh et~al\mbox{.}(2005)]%
        {hindmarsh2005Sundials}
\bibfield{author}{\bibinfo{person}{Alan~C. Hindmarsh},
  \bibinfo{person}{Peter~N. Brown}, \bibinfo{person}{Keith~E. Grant},
  \bibinfo{person}{Steven~L. Lee}, \bibinfo{person}{Radu Serban},
  \bibinfo{person}{Dan~E. Shumaker}, {and} \bibinfo{person}{Carol~S.
  Woodward}.} \bibinfo{year}{2005}\natexlab{}.
\newblock \showarticletitle{{SUNDIALS: Suite of nonlinear and
  differential/algebraic equation solvers}}.
\newblock  \bibinfo{volume}{31}, \bibinfo{number}{3} (\bibinfo{year}{2005}),
  \bibinfo{pages}{363--396}.
\newblock


\bibitem[Intel(2021)]%
        {MKL}
\bibfield{author}{\bibinfo{person}{Intel}.} \bibinfo{year}{2021}\natexlab{}.
\newblock \bibinfo{title}{Intel oneAPI Math Kernel Library}.
\newblock
  \bibinfo{howpublished}{\url{https://software.intel.com/content/www/us/en/develop/tools/oneapi/components/onemkl.html}}.
\newblock
\newblock
\shownote{Accessed: 2021-08-24}.


\bibitem[Intel(2023)]%
        {oneAPIcompilers}
\bibfield{author}{\bibinfo{person}{Intel}.} \bibinfo{year}{2023}\natexlab{}.
\newblock \bibinfo{title}{oneAPI DPC++ compiler}.
\newblock \bibinfo{howpublished}{\url{https://github.com/intel/llvm}}.
\newblock


\bibitem[{Intel Corp.}(2023)]%
        {oneapi2023}
\bibfield{author}{\bibinfo{person}{{Intel Corp.}}}
  \bibinfo{year}{2023}\natexlab{}.
\newblock \bibinfo{title}{oneAPI GPU Optimization Guide}.
\newblock
\newblock
\urldef\tempurl%
\url{https://www.intel.com/content/www/us/en/develop/documentation/oneapi-gpu-optimization-guide/top.html}
\showURL{%
\tempurl}
\newblock
\shownote{Accessed: Aug 2023}.


\bibitem[Joó et~al\mbox{.}(2019)]%
        {8945798}
\bibfield{author}{\bibinfo{person}{Bálint Joó}, \bibinfo{person}{Thorsten
  Kurth}, \bibinfo{person}{M.~A. Clark}, \bibinfo{person}{Jeongnim Kim},
  \bibinfo{person}{Christian~Robert Trott}, \bibinfo{person}{Dan Ibanez},
  \bibinfo{person}{Daniel Sunderland}, {and} \bibinfo{person}{Jack Deslippe}.}
  \bibinfo{year}{2019}\natexlab{}.
\newblock \showarticletitle{Performance Portability of a Wilson Dslash Stencil
  Operator Mini-App Using Kokkos and SYCL}. In \bibinfo{booktitle}{\emph{2019
  IEEE/ACM International Workshop on Performance, Portability and Productivity
  in HPC (P3HPC)}}. \bibinfo{pages}{14--25}.
\newblock
\urldef\tempurl%
\url{https://doi.org/10.1109/P3HPC49587.2019.00007}
\showDOI{\tempurl}


\bibitem[Kashi et~al\mbox{.}(2022)]%
        {kashiBatchedSparseIterative2022}
\bibfield{author}{\bibinfo{person}{Aditya Kashi}, \bibinfo{person}{Pratik
  Nayak}, \bibinfo{person}{Dhruva Kulkarni}, \bibinfo{person}{Aaron
  Scheinberg}, \bibinfo{person}{Paul Lin}, {and} \bibinfo{person}{Hartwig
  Anzt}.} \bibinfo{year}{2022}\natexlab{}.
\newblock \showarticletitle{Batched Sparse Iterative Solvers on {{GPU}} for the
  Collision Operator for Fusion Plasma Simulations}. In
  \bibinfo{booktitle}{\emph{2022 {{IEEE International Parallel}} and
  {{Distributed Processing Symposium}} ({{IPDPS}})}}.
  \bibinfo{pages}{157--167}.
\newblock
\showISSN{1530-2075}
\urldef\tempurl%
\url{https://doi.org/10.1109/IPDPS53621.2022.00024}
\showDOI{\tempurl}


\bibitem[Liegeois et~al\mbox{.}(2023a)]%
        {Kim2023}
\bibfield{author}{\bibinfo{person}{Kim Liegeois}, \bibinfo{person}{Sivasankaran
  Rajamanickam}, {and} \bibinfo{person}{Luc Berger-Vergiat}.}
  \bibinfo{year}{2023}\natexlab{a}.
\newblock \showarticletitle{Performance Portable Batched Sparse Linear
  Solvers}.
\newblock \bibinfo{journal}{\emph{IEEE Transactions on Parallel and Distributed
  Systems}} \bibinfo{volume}{34}, \bibinfo{number}{5} (\bibinfo{year}{2023}),
  \bibinfo{pages}{1524--1535}.
\newblock
\urldef\tempurl%
\url{https://doi.org/10.1109/TPDS.2023.3249110}
\showDOI{\tempurl}


\bibitem[Liegeois et~al\mbox{.}(2023b)]%
        {liegeoisPerformancePortableBatched2023}
\bibfield{author}{\bibinfo{person}{Kim Liegeois}, \bibinfo{person}{Sivasankaran
  Rajamanickam}, {and} \bibinfo{person}{Luc {Berger-Vergiat}}.}
  \bibinfo{year}{2023}\natexlab{b}.
\newblock \showarticletitle{Performance {{Portable Batched Sparse Linear
  Solvers}}}.
\newblock \bibinfo{journal}{\emph{IEEE Transactions on Parallel and Distributed
  Systems}} \bibinfo{volume}{34}, \bibinfo{number}{5} (\bibinfo{date}{May}
  \bibinfo{year}{2023}), \bibinfo{pages}{1524--1535}.
\newblock
\showISSN{1558-2183}
\urldef\tempurl%
\url{https://doi.org/10.1109/TPDS.2023.3249110}
\showDOI{\tempurl}


\bibitem[Nguyen et~al\mbox{.}(2023)]%
        {nguyenReproducibilityArtifactGinkgo2023}
\bibfield{author}{\bibinfo{person}{Phuong Nguyen}, \bibinfo{person}{Pratik
  Nayak}, {and} \bibinfo{person}{Hartwig Anzt}.}
  \bibinfo{year}{2023}\natexlab{}.
\newblock \bibinfo{title}{Reproducibility Artifact for {{Ginkgo}}'s Batched
  Iterative Solvers for {{GPUs}} with {{CUDA}}, {{HIP}} and {{SYCL}}
  Programming Models}.
\newblock \bibinfo{howpublished}{Zenodo}.
\newblock
\urldef\tempurl%
\url{https://doi.org/10.5281/ZENODO.8247538}
\showDOI{\tempurl}


\bibitem[Nickolls et~al\mbox{.}(2008)]%
        {10.1145/1401132.1401152}
\bibfield{author}{\bibinfo{person}{John Nickolls}, \bibinfo{person}{Ian Buck},
  \bibinfo{person}{Michael Garland}, {and} \bibinfo{person}{Kevin Skadron}.}
  \bibinfo{year}{2008}\natexlab{}.
\newblock \showarticletitle{Scalable Parallel Programming with CUDA}. In
  \bibinfo{booktitle}{\emph{ACM SIGGRAPH 2008 Classes}} (Los Angeles,
  California) \emph{(\bibinfo{series}{SIGGRAPH '08})}.
  \bibinfo{publisher}{Association for Computing Machinery},
  \bibinfo{address}{New York, NY, USA}, Article \bibinfo{articleno}{16},
  \bibinfo{numpages}{14}~pages.
\newblock
\showISBNx{9781450378451}
\urldef\tempurl%
\url{https://doi.org/10.1145/1401132.1401152}
\showDOI{\tempurl}


\bibitem[NVIDIA(2021)]%
        {cuBLAS}
\bibfield{author}{\bibinfo{person}{NVIDIA}.} \bibinfo{year}{2021}\natexlab{}.
\newblock \bibinfo{title}{{cuBLAS} - Basic linear algebra on {NVIDIA} {GPU}s}.
\newblock \bibinfo{howpublished}{\url{https://developer.nvidia.com/cublas}}.
\newblock
\newblock
\shownote{Accessed: 2021-08-24}.


\bibitem[Ohshima et~al\mbox{.}(2019)]%
        {batchedgemv}
\bibfield{author}{\bibinfo{person}{Satoshi Ohshima}, \bibinfo{person}{Ichitaro
  Yamazaki}, \bibinfo{person}{Akihiro Ida}, {and} \bibinfo{person}{Rio
  Yokota}.} \bibinfo{year}{2019}\natexlab{}.
\newblock \showarticletitle{Optimization of Numerous Small
  Dense-Matrix–Vector Multiplications in H-Matrix Arithmetic on {GPU}}. In
  \bibinfo{booktitle}{\emph{2019 IEEE 13th International Symposium on Embedded
  Multicore/Many-core Systems-on-Chip (MCSoC)}}. \bibinfo{pages}{9--16}.
\newblock
\urldef\tempurl%
\url{https://doi.org/10.1109/MCSoC.2019.00009}
\showDOI{\tempurl}


\bibitem[Reinders et~al\mbox{.}(2021)]%
        {dpcpp2021}
\bibfield{author}{\bibinfo{person}{James Reinders}, \bibinfo{person}{Ben
  Ashbaugh}, \bibinfo{person}{James Brodman}, \bibinfo{person}{Michael
  Kinsner}, \bibinfo{person}{John Pennycook}, {and} \bibinfo{person}{Xinmin
  Tian}.} \bibinfo{year}{2021}\natexlab{}.
\newblock \bibinfo{booktitle}{\emph{Data Parallel C++: Mastering DPC++ for
  Programming of Heterogeneous Systems using C++ and SYCL}}.
\newblock
\showISBNx{978-1-4842-5573-5}
\urldef\tempurl%
\url{https://doi.org/10.1007/978-1-4842-5574-2}
\showDOI{\tempurl}


\bibitem[Reyes et~al\mbox{.}(2020)]%
        {10.1145/3388333.3388649}
\bibfield{author}{\bibinfo{person}{Ruyman Reyes}, \bibinfo{person}{Gordon
  Brown}, \bibinfo{person}{Rod Burns}, {and} \bibinfo{person}{Michael Wong}.}
  \bibinfo{year}{2020}\natexlab{}.
\newblock \showarticletitle{SYCL 2020: More than Meets the Eye}. In
  \bibinfo{booktitle}{\emph{Proceedings of the International Workshop on
  OpenCL}} (Munich, Germany) \emph{(\bibinfo{series}{IWOCL '20})}.
  \bibinfo{publisher}{Association for Computing Machinery},
  \bibinfo{address}{New York, NY, USA}, Article \bibinfo{articleno}{4},
  \bibinfo{numpages}{1}~pages.
\newblock
\showISBNx{9781450375313}
\urldef\tempurl%
\url{https://doi.org/10.1145/3388333.3388649}
\showDOI{\tempurl}


\bibitem[Saad(2003)]%
        {saadIterativeMethodsSparse2003}
\bibfield{author}{\bibinfo{person}{Yousef Saad}.}
  \bibinfo{year}{2003}\natexlab{}.
\newblock \bibinfo{booktitle}{\emph{Iterative {Methods} for {Sparse} {Linear}
  {Systems}} (\bibinfo{edition}{second} ed.)}.
\newblock \bibinfo{publisher}{Society for Industrial and Applied Mathematics}.
\newblock
\showISBNx{978-0-89871-534-7 978-0-89871-800-3}
\urldef\tempurl%
\url{https://doi.org/10.1137/1.9780898718003}
\showDOI{\tempurl}


\bibitem[Smolarski et~al\mbox{.}(2022)]%
        {9912708}
\bibfield{author}{\bibinfo{person}{Dennis~C. Smolarski},
  \bibinfo{person}{F.~Douglas Swesty}, {and} \bibinfo{person}{Alan~C. Calder}.}
  \bibinfo{year}{2022}\natexlab{}.
\newblock \showarticletitle{Performance of an Astrophysical Radiation
  Hydrodynamics Code under Scalable Vector Extension Optimization}. In
  \bibinfo{booktitle}{\emph{2022 IEEE International Conference on Cluster
  Computing (CLUSTER)}}. \bibinfo{pages}{545--548}.
\newblock
\urldef\tempurl%
\url{https://doi.org/10.1109/CLUSTER51413.2022.00071}
\showDOI{\tempurl}


\bibitem[Valero-Lara et~al\mbox{.}(2018)]%
        {cuThomasBatch}
\bibfield{author}{\bibinfo{person}{Pedro Valero-Lara}, \bibinfo{person}{I.
  Mart{\'i}nez-P{\'e}rez}, \bibinfo{person}{Ra{\"u}l Sirvent},
  \bibinfo{person}{X. Martorell}, {and} \bibinfo{person}{Antonio~J. Pe{\~n}a}.}
  \bibinfo{year}{2018}\natexlab{}.
\newblock \showarticletitle{cu{T}homas{B}atch and {cuThomasVBatch}, {CUDA}
  routines to compute batch of tridiagonal systems on {NVIDIA GPUs}}.
\newblock \bibinfo{journal}{\emph{Concurrency and Computation: Practice and
  Experience}}  \bibinfo{volume}{30} (\bibinfo{year}{2018}).
\newblock


\end{thebibliography}


\end{document}